\documentclass[twocolumn]{aastex63}
\usepackage{amsmath}
\usepackage{textcomp}
\received{June 1, 2019}
\revised{January 10, 2019}
\accepted{\today}

\shortauthors{Li et al.}

\newcommand{\hii}{\ion{H}{2}}
\newcommand{\oiii}{[\ion{O}{3}]$_{\rm 88 \mu m}$}
\newcommand{\oi}{[\ion{O}{1}]$_{\rm 146 \mu m}$}
\newcommand{\oione}{[\ion{O}{1}]$_{\rm 63 \mu m}$}
\newcommand{\nii}{[\ion{N}{2}]$_{\rm 122 \mu m}$}

\newcommand{\niis}{[\ion{N}{2}]$_{\rm 205 \mu m}$}
\newcommand{\cii}{[\ion{C}{2}]$_{\rm 158 \mu m}$}
\newcommand{\lsun}{$\rm L_{\sun}$}
\newcommand{\msun}{$\rm M_{\sun}$}
\newcommand{\nele}{$n_{e}$}
\newcommand{\pcubcm}{$\rm cm^{-3}$}
\newcommand{\kmps}{$\rm km\ s^{-1}$}
\newcommand{\qso}{J2310+1855}
\newcommand{\qsoz}{$z = 6.003$}

\begin{document}

\shortauthors{Li et al.}
%\shorttitle{}
\title{Ionized and atomic interstellar medium in the \qsoz{} quasar SDSS J2310+1855}

\correspondingauthor{Jianan Li}
\email{jiananl@pku.edu.cn}
\correspondingauthor{Ran Wang}
\email{rwangkiaa@pku.edu.cn}

\author{Jianan Li}
\affiliation{Department of Astronomy, School of Physics, Peking University, Beijing 100871, China}
\affiliation{Kavli Institute for Astronomy and Astrophysics, Peking University, Beijing 100871, China}

\author{Ran Wang}
\affiliation{Kavli Institute for Astronomy and Astrophysics, Peking University, Beijing 100871, China}

\author{Pierre Cox}
\affiliation{Institut d'Astrophysique de Paris, Sorbonne Universit\'{e}, CNRS, UMR 7095, 98 bis bd Arago, 75014 Paris, France}

\author{Yu Gao}
\affiliation{Purple Mountain Observatory $\&$ Key Laboratory for Radio Astronomy, Chinese Academy of Sciences, 10 Yuanhua Road, Nanjing
210033, PR China}
\affiliation{Department of Astronomy, Xiamen University, Xiamen, Fujian 361005, China}

\author{Fabian Walter}
\affiliation{Max-Planck-Institut for Astronomie, K\"{o}nigstuhl 17, D-69117 Heidelberg, Germany}

\author{Jeff Wagg}
\affiliation{SKA Organization, Lower Withington Macclesfield, Cheshire SK11 9DL, UK}

\author{Karl M. Menten}
\affiliation{Max-Planck-Institut f{\"u}r Radioastronomie, Auf dem H\"{u}gel 69, 53121 Bonn, Germany}
\author{Frank Bertoldi}
\author{Yali Shao}
\affiliation{Max-Planck-Institut f{\"u}r Radioastronomie, Auf dem H\"{u}gel 69, 53121 Bonn, Germany}

\affiliation{Argelander-Institut f{\"u}r Astronomie, University at Bonn, Auf dem H\"{u}gel 71, D-53121 Bonn, Germany}
\author{Bram P. Venemans}
\affiliation{Max-Planck-Institut for Astronomie, K\"{o}nigstuhl 17, D-69117 Heidelberg, Germany}
\author{Roberto Decarli}
\affiliation{INAF -- Osservatorio di Astrofisica e Scienza dello Spazio, via Gobetti 93/3, 40129 Bologna, Italy}

\author{Dominik Riechers}
\affiliation{Department of Astronomy, Cornell University, Space Sciences Building, Ithaca, NY 14853, USA}
\affiliation{Max-Planck-Institut for Astronomie, K\"{o}nigstuhl 17, D-69117 Heidelberg, Germany}

\author{Roberto Neri}
\affiliation{Institut de Radioastronomie Millimétrique, Saint Martin d'Hères, F-38406, France}

\author{Xiaohui Fan}
\affiliation{Steward Observatory, University of Arizona, 933 North Cherry Avenue, Tucson, AZ 85721, USA}

%\author{Chris L. carilli}
%\affiliation{Cavendish Laboratory, 19 J. J. Thomson Avenue, Cambridge CB3 0HE, UK}
%\affiliation{National Radio Astronomy Observatory, Socorro, NM 87801-0387, USA}

\author{Alain Omont}
\affiliation{Institut d'Astrophysique de Paris, Sorbonne Universit\'{e}, CNRS, UMR 7095, 98 bis bd Arago, 75014 Paris, France}

\author{Desika Narayanan}
\affiliation{Department of Astronomy, University of Florida, 211 Bryant Space Science Center, Gainesville, FL 32611, USA}

\begin{abstract}
Observing the interstellar medium (ISM) in $z \gtrsim 6$ quasar host galaxies is essential for understanding the coevolution between the supermassive black holes and their hosts. To probe the gas physical conditions and search for imprints of active galactic nuclei (AGN) on the ISM, we report ALMA observations of the \nii{} and \oi{} lines and the underlying continuum from the \qsoz{} quasar SDSS J231038.88+185519.7. Together with previous \cii{} and \oiii{} observations, we use the ratios of these fine-structure lines to probe the ISM properties. Similar to other high-$z$ systems, this object exhibits a \cii{}/\oi{} ratio comparable to the lowest values found in local (Ultra)luminous infrared galaxies, suggesting a ``warmer'' and ``denser'' gas component compared to typical local systems.  The \oiii{}/\oi{}  ratio is  lower than that of other local and  high-$z$ systems, indicating a smaller ionized gas fraction in this quasar. 
The \oiii{}/\nii{} ratio is comparable to that of local systems, and suggests a metallicity of $Z/Z_{\sun}$=1.5$-$2.1. 
Based on the \nii{} detection, we estimate that $17\%$ of the \cii{} emission is associated with ionized gas.
The \nii{} line shows a ``flux deficit'' comparable to local systems.  The \oi{} line, with a \oi{}/FIR ratio $\textgreater 2\times$ than expected from the local relation, indicates no \oi{} deficit. The low \cii{}/\oi{} ratio, together with the high \oi{}/FIR ratio in \qso{}, reveals that the warm and dense gas is likely a result of AGN heating of the ISM.
%reveals that the ISM is likely heated by the AGN
%The \oi{} line, with a \oi{}/FIR ratio $\textgreater 2\times$ that expected from the local relation, indicates no \oi{} deficit, revealing warm and dense gas possibly heated by the AGN.

\end{abstract}

%% Keywords should appear after the \end{abstract} command. 
%% See the online documentation for the full list of available subject
%% keywords and the rules for their use.
%\keywords{/////////}

 \section{Introduction}
Observations of the interstellar medium (ISM) in the host galaxies of high-$z$ quasars reveal key aspects in the processes of galaxy formation and evolution as well as the growth of supermassive black holes (SMBHs). %, which is essential for our understanding of the coevolution between SMBHs and galaxies.
%A sample of $z \gtrsim 6$ quasars, which are hosting SMBHs with masses of $\sim 10^{9}$ \msun{}, accreting gas at rates near the Eddington limit, are embedded in gas rich host galaxies with extremely high star formation rates, e.g., $\sim 10^{3}$ $\rm M_{\sun}\ yr^{-1}$ (e.g., \citealt{wang11,wang13,wang16,wang19}; \citealt{carilli13};  \citealt{shao17,shao19}; \citealt{venemans17a,venemans17b}; \citealt{decarli18}; \citealt{li20}). 
A sample of optically luminous quasars was discovered at $z \gtrsim 6$, which are hosting SMBHs with masses up to $\sim 10^{10}$ \msun{} and accreting close to the Eddington limit (e.g., \citealt{jiang07}; \citealt{wu15}; \citealt{shen19}). Some of these objects are also detected in bright dust continuum and line emission at submillimeter/millimeter wavelengths, suggesting they are embedded in gas-rich host galaxies with extremely high star formation rates, e.g., $\sim 10^{3}$ $\rm M_{\sun}\ yr^{-1}$ (e.g., \citealt{wang11,wang13,wang16,wang19}; \citealt{carilli13};  \citealt{shao17,shao19}; \citealt{venemans17a,venemans17b}; \citealt{decarli18}; \citealt{li20}). 
These millimeter bright quasars are  among the most extreme systems in the early universe and serve as excellent candidates to study the coevolution of the SMBHs and their host galaxies at the earliest evolutionary stages.
During the past few decades, the most sensitive submillimeter to millimeter wavelength interferometers, e.g., ALMA and NOEMA, have demonstrated their extreme power in detecting the main coolants in photo-dissociation regions (PDRs) from quasar host galaxies at $z \gtrsim 6$, e.g., the \oi{} and \cii{} lines from the neutral ISM (e.g., \citealt{wang13, wang16, wang19}; \citealt{shao17}; \citealt{venemans17a,venemans17b}; \citealt{decarli18}; \citealt{novak19}; \citealt{yang19}), the CO and water ($\rm H_{2}O$) lines from the molecular ISM (e.g., \citealt{riechers09}; \citealt{gallerani14};  \citealt{shao19};  \citealt{wangf19}; \citealt{yang19}; \citealt{li20}), and the \oiii{} and \nii{} lines from the ionized ISM (e.g., \citealt{walter18}; \citealt{hashimoto18}; \citealt{novak19}).
These far-infrared (FIR) fine-structure (FS) lines, together with the molecular lines, serve as diagnostics of the physical and chemical conditions  (e.g., the temperature, density, radiation field strength, and metallicity) of the ISM in different phases, and allow us to discriminate between different gas-heating mechanisms (e.g., ultraviolet photons and X-rays in PDRs and X-ray dominated regions (XDRs), respectively, cosmic rays, and shocks).

The \cii{} FS line, which is typically the brightest FIR emission line and the main coolant of the ISM, has long been the workhorse to probe the neutral gas content in $z \gtrsim 6$ quasars (e.g., \citealt{wang13,wang16,wang19}; \citealt{shao17}; \citealt{venemans17a,venemans17b}; \citealt{decarli18}). 
With an ionization potential (11.26 eV) lower than that of hydrogen, the \cii{} emission can coexist in both the ionized and neutral gas.
Recent studies of local (U)LIRGs, active galactic nuclei (AGNs), and high-redshift quasars reveal that most of the \cii{}  emission originates in the neutral gas phase ($\gtrsim 70\%$; e.g., \citealt{herrera16,herrera18a}; \citealt{diaz17}; \citealt{novak19}).
The molecular CO and water ($\rm H_{2}O$) lines, being usually over an order of magnitude fainter than the \cii{} line, represent the brightest molecular ISM emissions in  $z \gtrsim 6$ quasars. 
Recently, a handful of other FS lines from the ionized and atomic ISM, e.g., \oi{}, \oiii{}, and \nii{} tracing the atomic and ionized ISM were detected in the quasar host galaxies and star-forming galaxies $z \gtrsim 6$ (e.g.,  \citealt{walter18}; \citealt{hashimoto18}; \citealt{novak19}; \citealt{yang19}). 
In particular, the recent tentative detection of the \oione{} line with a luminosity four times that of the \cii{} line in a $z=6.023$ lensed dusty star forming galaxy G09.83808, and the \oi{} detected in the $z=6.52$ quasar J0439+1634 with a \oi{}/\cii{} ratio of 0.3, make the \oione{} and \oi{} lines excellent tracers of the neutral gas at $z\gtrsim 6$ in addition to \cii{} (\citealt{yang19}; \citealt{rybak20}).
Combinations of different FS lines provide rich information about the multiphase ISM, e.g., the \nii{}/\niis{} ratio as a probe of ionized gas density (e.g.,  \citealt{oberst06,oberst11}; \citealt{novak19}),  the \cii{}/\niis{} ratio as an indicator of  the \cii{} origin (e.g., \citealt{pavesi16,pavesi19}; \citealt{herrera18b}; \citealt{debreuck19}), the \cii{}/\oione{} (or \oi{}) ratio as a tracer of neutral gas density and temperature (e.g.,  \citealt{oberst06,oberst11}; \citealt{diaz17}; \citealt{herrera18a}), and the \nii{}/\oiii{} ratio as a diagnostic of the ionization parameter and metallicity (e.g., \citealt{nagao11}; \citealt{pereira17}; \citealt{rigopoulou18}). 
In addition, the FS lines also serve as tracers of the star formation rate (SFR) and infrared luminosity in (ultra)luminous infrared galaxies ((U)LIRGs), and studying the ratio of FS line and FIR luminosity addresses the origin of the FIR line deficit, i.e., a decrease in the line to FIR luminosity ratio for many of the most luminous objects, e.g., local (U)LIRGs and AGNs (e.g., \citealt{farrah13}; \citealt{delooze14}; \citealt{herrera18a}).

In this work, we present an ISM study of the $z=6.003$ quasar SDSS J231038.88+185519.7 (hereafter J2310+1855). 
Being one of the FIR brightest quasars at $z \gtrsim 6$, J2310+1855 has been widely studied in lines probing different phases of the ISM, e.g., \cii{} emission from the neutral gas; the CO, $\rm H_{2}O$, and $\rm OH^{+}$ lines from the molecular material; and the  \oiii{} line from the ionized gas (\citealt{wang13}; \citealt{hashimoto18}; \citealt{shao19}; \citealt{li20}).
In a previous paper \citep{li20}, we report a study of the CO spectral line energy distribution (SLED) of this object. We detected highly excited molecular gas with complex excitation mechanisms in which heating from the powerful quasar could be involved.
In this work, we report new ALMA observations of \nii{} and \oi{} FS lines in J2310+1855, to further probe the impact of the AGN and as well as the star formation from its host galaxy, on the multiphase ISM of this quasar.
%Our recent work presenting its extensive CO spectral line energy distribution (SLED), suggests several possible mechanisms how a powerful quasar might heat its molecular gas \citep{li20}. It is thus essential to check, through observations of ISM emission lines that probe other constituents of its ISM, whether the powerful quasar contributes to the gas heating. We report new ALMA observations of \nii{} and \oi{} FS lines in J2310+1855, to further probe impact of the AGN on the one hand and of vigorous star formation of its host galaxy on the other on the multi-phase ISM of this quasar.
%probe the respective impacts on the multi-phase ISM of the AGN and the starbursting host galaxy of this $z\approx6$ quasar.  
%searching for any evidence that the AGN and the nearby starburst activity in the host galaxy might impact on the multi-phase ISM. 
We adopt a standard $\Lambda$CDM cosmology with $H_{0}=70\ \rm km \ s^{-1}\ Mpc^{-1}$ and $\Omega_{\rm m}=0.3$. 
With the adopted cosmological parameters, 1$''$ corresponds to 5.7 kpc at the redshift of J2310+1855 and the luminosity distance to the quasar is 57.8 Gpc.

\section{Observations}
We observed the [\ion{O}{1}]$\rm\ 146 \mu m$ ($\nu _{\rm rest}$= 2060.0690 GHz) and [\ion{N}{2}]$\rm\ 122 \mu m$ ($\nu _{\rm rest}$= 2459.3801 GHz) FS lines from the quasar \qso{} at \qsoz{} with ALMA. The observations were performed as part of Cycle 3 program ID 2015.1.01265.S.
%We observed the \oi{} and  \nii{} fine structural lines between 2016 April 9 and November 22 at a beam size of $\sim 0\farcs{5}-0\farcs{6}$ in ALMA band 7. 
The \oi{} line was observed on 2016 April 14 and November 22 with a total observing time of 73.4 minutes on source. The array configuration was C36-3, where the number of antennas used was 36--38 and the baseline lengths were between 15 and 539 meters. 
We set one of the four 1.875 GHz spectral windows centered on the redshifted \oi{} line at 294.1653 GHz, and the remaining three spectral windows observed the continuum emission.
The bandpass and flux calibrator was J2253+1608. J2300+1655 was used for phase calibration.
On 2016 April 9 and June 16, we observed the \nii{}  line in the C36-2 configuration with a total observing time of 67.8 minutes on source.
The number of antennas was 38--44 and the baseline lengths were between 15 and 377 m. We tuned one  1.875 GHz width spectral window centered at the redshifted \nii{} frequency, 351.1845 GHz, to observe the line, and the other three 1.875 GHz width spectral windows to observe the continuum emission. The bandpass and phase calibrator was J2253+1608, and the flux calibrator was Pallas.

We reduced the ALMA data with the Common Astronomy Software Applications (CASA) software package version 4.7.0 \citep{casa07}, following the standard ALMA pipeline. 
We imaged the data using the CLEAN task in CASA with natural weighting.  The synthesized beam sizes were $0\farcs56$ $\times$ $0\farcs46$ and $0\farcs47$ $\times$ $0\farcs39$ at the observing frequency of the \nii{} and \oi{} lines, respectively.
We generated the datacube for spectral line imaging by executing the UVCONTSUB task in CASA, through subtracting a first-order polynomial continuum from the original datacube.
We finally binned the \oi{} datacube to 16 MHz (16 km $\rm s^{-1}$) wide channels, resulting in a rms of 0.20  $\rm mJy\ beam^{-1}$ per binned channel.
As for the \nii{} datacube, we binned them to 94 MHz (80 km $\rm s^{-1}$) wide channels, yielding a rms of 0.23  $\rm mJy\ beam^{-1}$ per binned channel.
%The sensitivity of the underlying continuum is 25 (49) $\rm \mu Jy\ beam^{-1}$ at 289.2 (344.2) GHz. 
By averaging the data over 281.2 (336.2) and 295.2 (352.2) GHz in the regions free of line emission, we find a continuum sensitivity of 25 (49) $\rm \mu Jy\ beam^{-1}$ at 289.2 (344.2) GHz from our observations.

\section{Results}
\begin{figure*}
\gridline{\fig{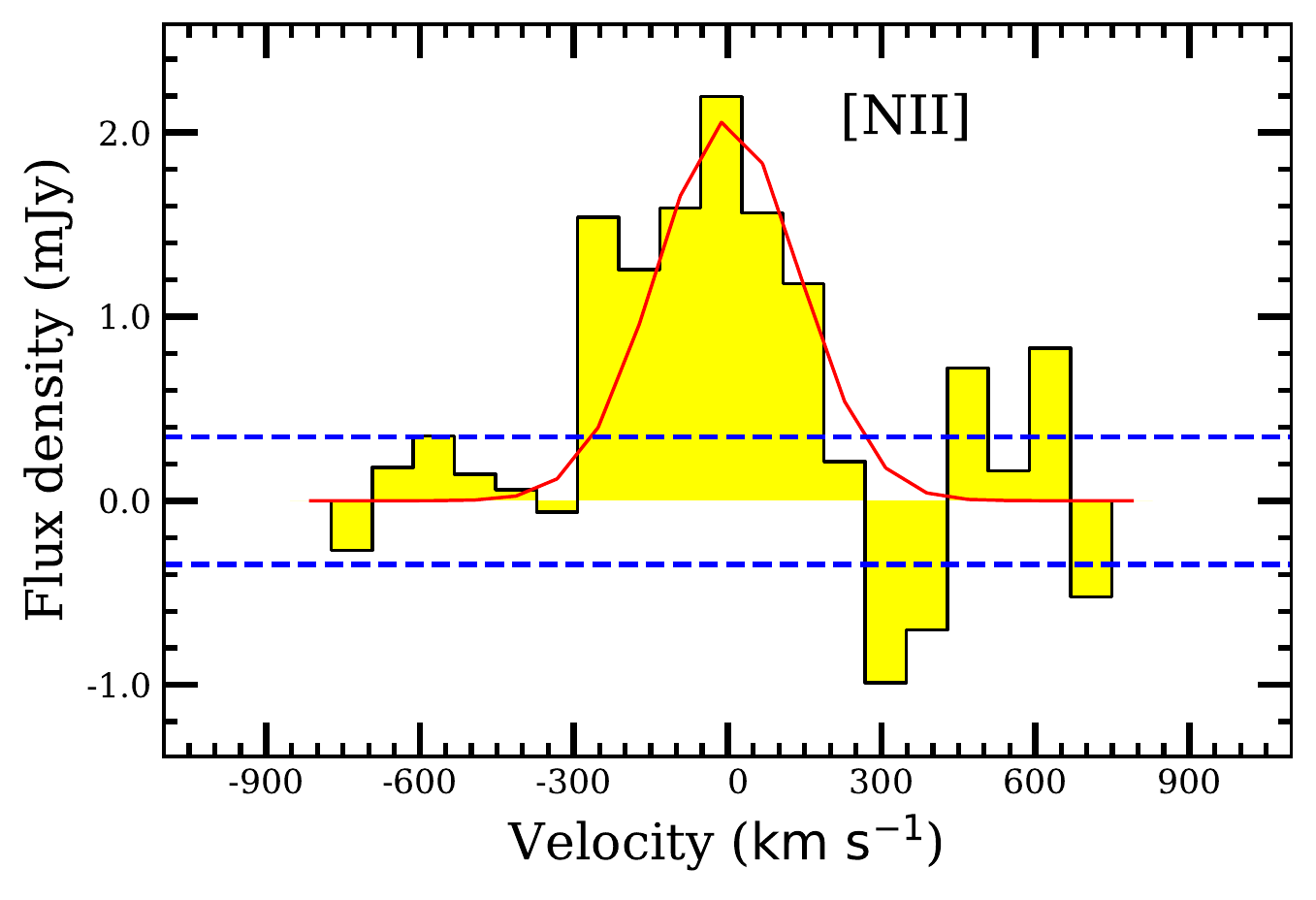}{0.499\textwidth}{(a)}
          \fig{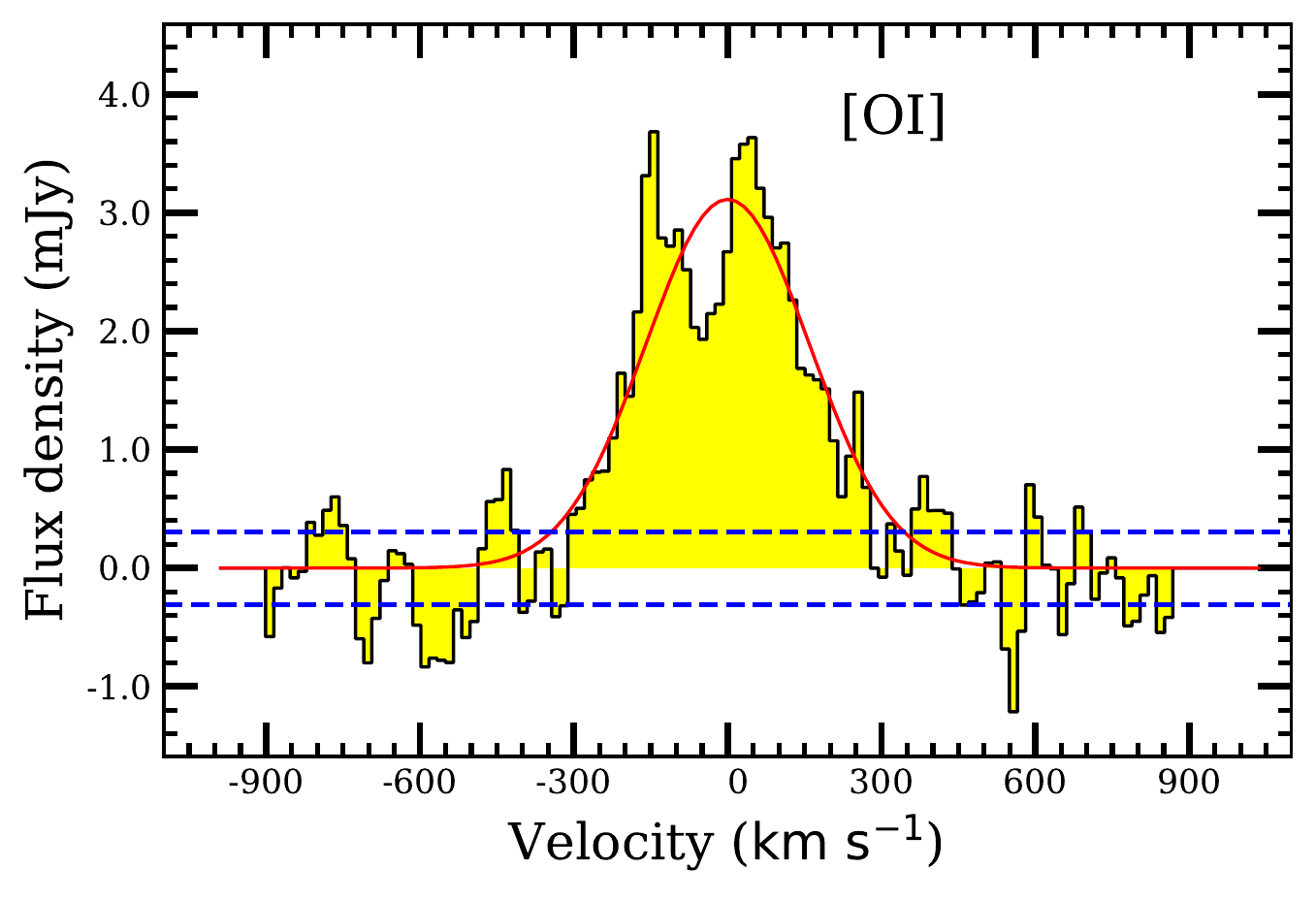}{0.499\textwidth}{(b)}
          }
          
\gridline{
          \fig{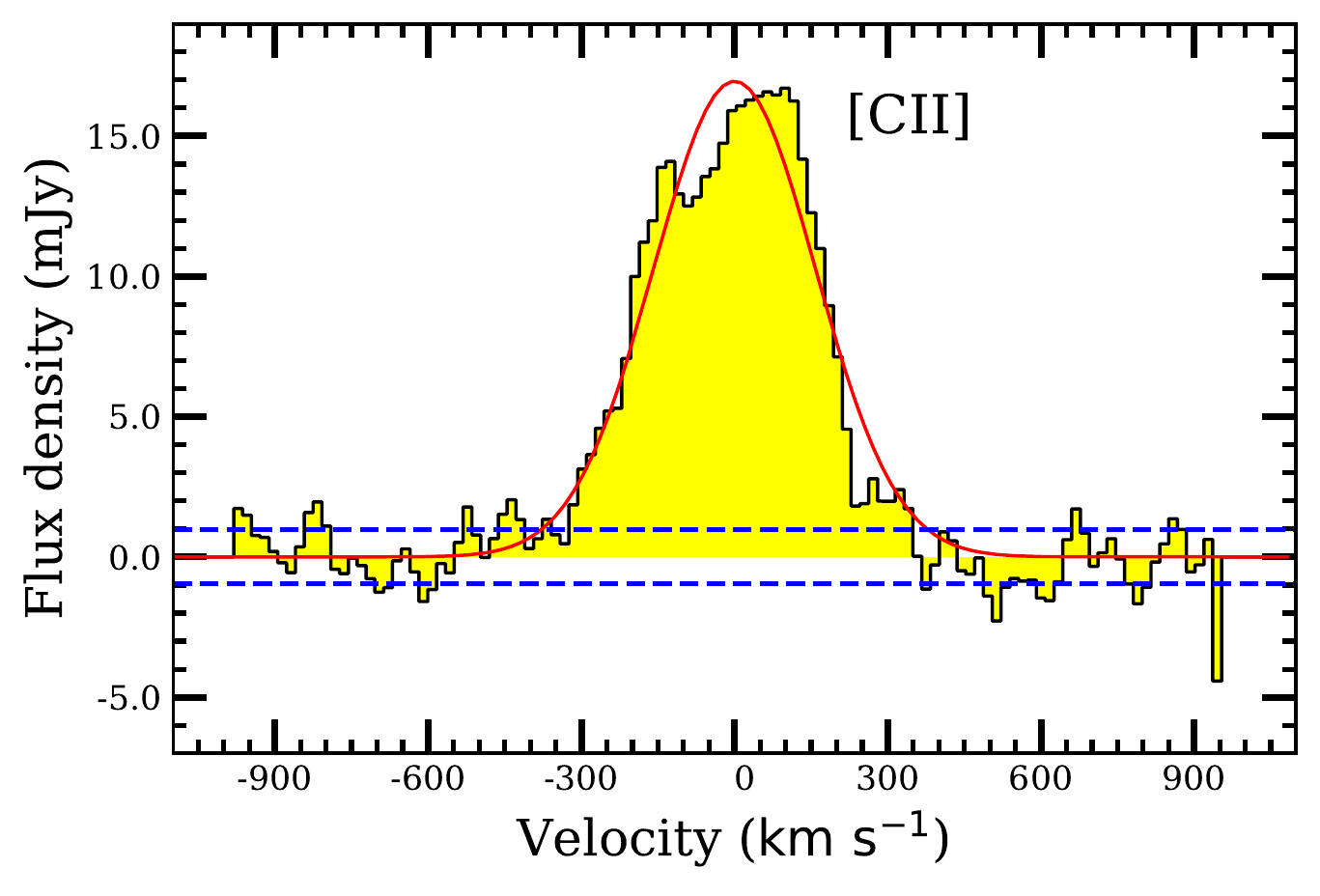}{0.499\textwidth}{(c)}
          \fig{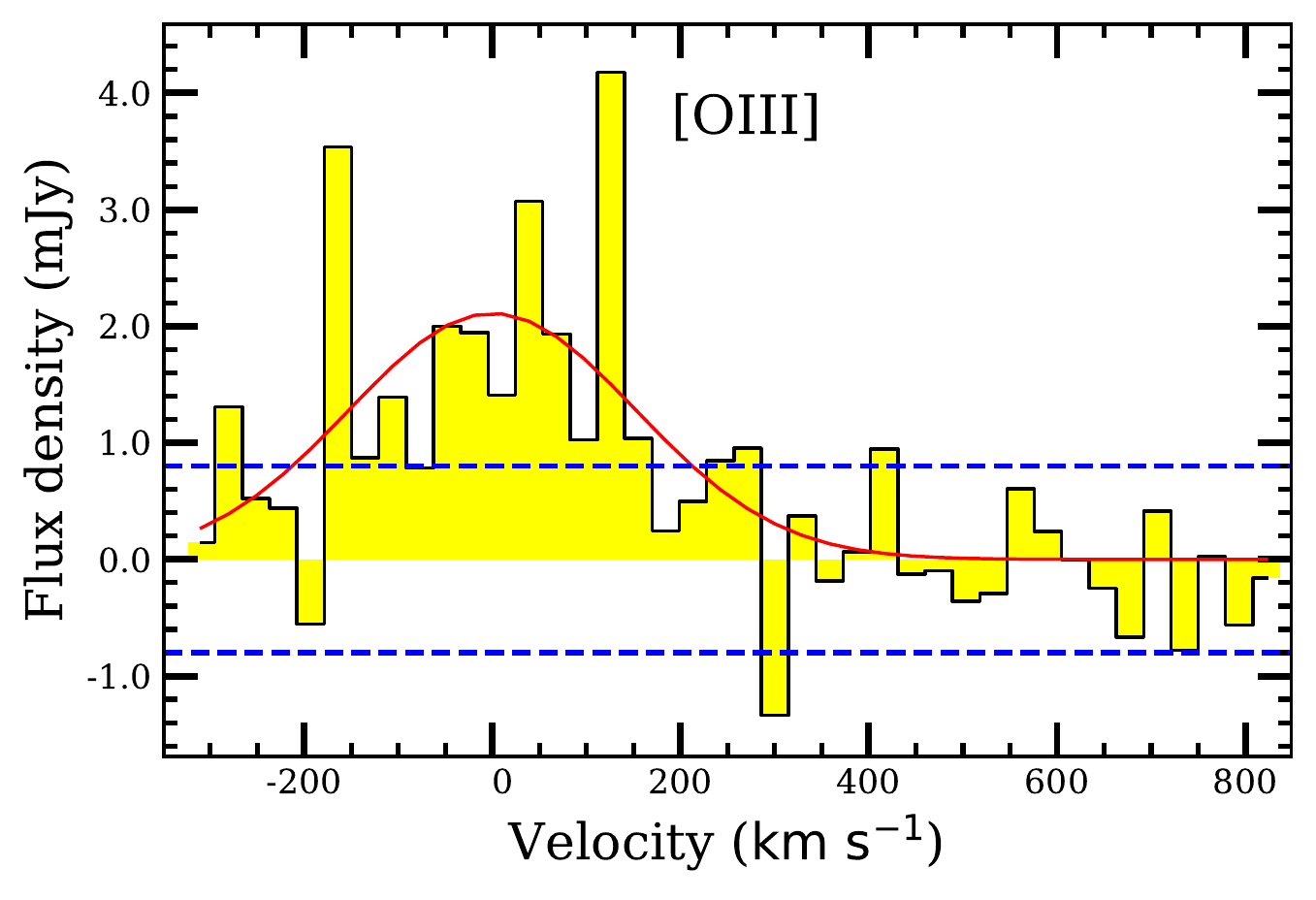}{0.499\textwidth}{(d)}
          }
%\gridline{\fig{RS_Oph.pdf}{0.3\textwidth}{(d)}
%          \fig{U_Sco.pdf}{0.3\textwidth}{(e)}
 %         }
%\gridline{\fig{radial_profile.pdf}{0.3\textwidth}{(f)}}
\caption{Spectra of \nii{} and  \oi{} (this work),  \cii{} \citep{wang13}, and \oiii{} \citep{hashimoto18} in the  \qsoz{} quasar \qso{}. Data are shown in yellow histograms. The red solid lines are Gaussian profile fits to the spectra with the line centers fixed to the \cii{} redshift of $z=6.0031$. The blue dashed lines indicate the $\pm$1$\sigma$ noise. The continuum has been subtracted for each of the spectra.
\label{spectra}}
\end{figure*}

Both the \oi{} and \nii{} lines and the underlying continuum are detected in our ALMA observations.
The spectra of the \oi{} and \nii{} lines, which are presented in Figure  \ref{spectra} together with the \cii{} spectrum from \citet{wang13} and that of the \oiii{} emission line from \citet{hashimoto18}, have been extracted within a circular aperture with a radius of $0 \farcs 8$ centered at the peak spaxel. For  the \oi{} and \nii{} lines, we measure the flux and width by fitting a single Gaussian profile to the spectrum while the line center is fixed to that of the \cii{} redshift \citep{wang13}. The measured fluxes are 0.72 $\pm$ 0.20 and 1.25 $\pm$ 0.08 Jy $\rm km\ s^{-1}$ for the \nii{}  and the \oi{} lines \footnote{We note that the line flux measured from clipping all the line emitting channels  is consistent with that from a single Gaussian profile fitting. Because of the absorption-like feature in the spectrum,  fitting a Gaussian profile to the spectrum  could overestimate the uncertainties.}, respectively. 
The resulting line widths of 328 $\pm$ 58 and 376 $\pm$ 16 km s$^{-1}$ in FWHM for \nii{} and \oi{}  are consistent with those measured for \cii{} (393 $\pm$ 21 km s$^{-1}$, \citealt{wang13}), \oiii{} (333 $\pm$ 72 km s$^{-1}$, \citealt{hashimoto18}), and the CO lines ($\sim$ 400 km s$^{-1}$, \citealt{li20}).
The measured line luminosity, flux, source size, and line width are listed in Table \ref{tabline}. 
The spectrum of \oi{} shows a clear absorption-like feature signal-to-noise ratio (S/N  $\textgreater \ 5$) in the blue part close to the line center frequency. Interestingly, this feature is also seen in the previous \cii{} data \citep{wang13}.
The \nii{} spectrum suggests a tentative dip at a similar velocity to that of the \cii{} and \oi{} emission lines, while the low-S/N spectrum of the \oiii{} line reveals no clear sign of an absorption-like feature.  Future observations, possibly at higher sensitivity are required to confirm whether this feature exists in the ionized gas tracers, e.g., \nii{} and [\ion{O}{3}]$_{\rm 88 \mu m}$.
%As for the \nii{} spectrum, we find a deficit at a similar velocity to that of the \cii{} and \oi{} lines. No absorption like feature is found in the \oiii{} spectrum with low S/N ratio \citep{hashimoto18}. 
However, this feature is not observed in the molecular CO lines, possibly resulting from the different gas components that the CO and FIR FS lines trace (\citealt{shao19}; \citealt{li20}).  
%To summarize, the absorption-like feature is not observed in the molecular gas, but is found in gas from neutral phase, e.g. \oi{} and \cii{} lines. 
%A suspicious absorption-like feature is found in the ionized gas tracer \nii{}, although future high sensitivity observations of the FIR FS lines from the ionized gas phase are needed to confirm whether this feature exists in the ionized gas tracers e.g., \nii{} and [\ion{O}{3}]$_{\rm 88 \mu m}$.

\begin{figure*}
\centering 
\includegraphics[width=1.\textwidth]{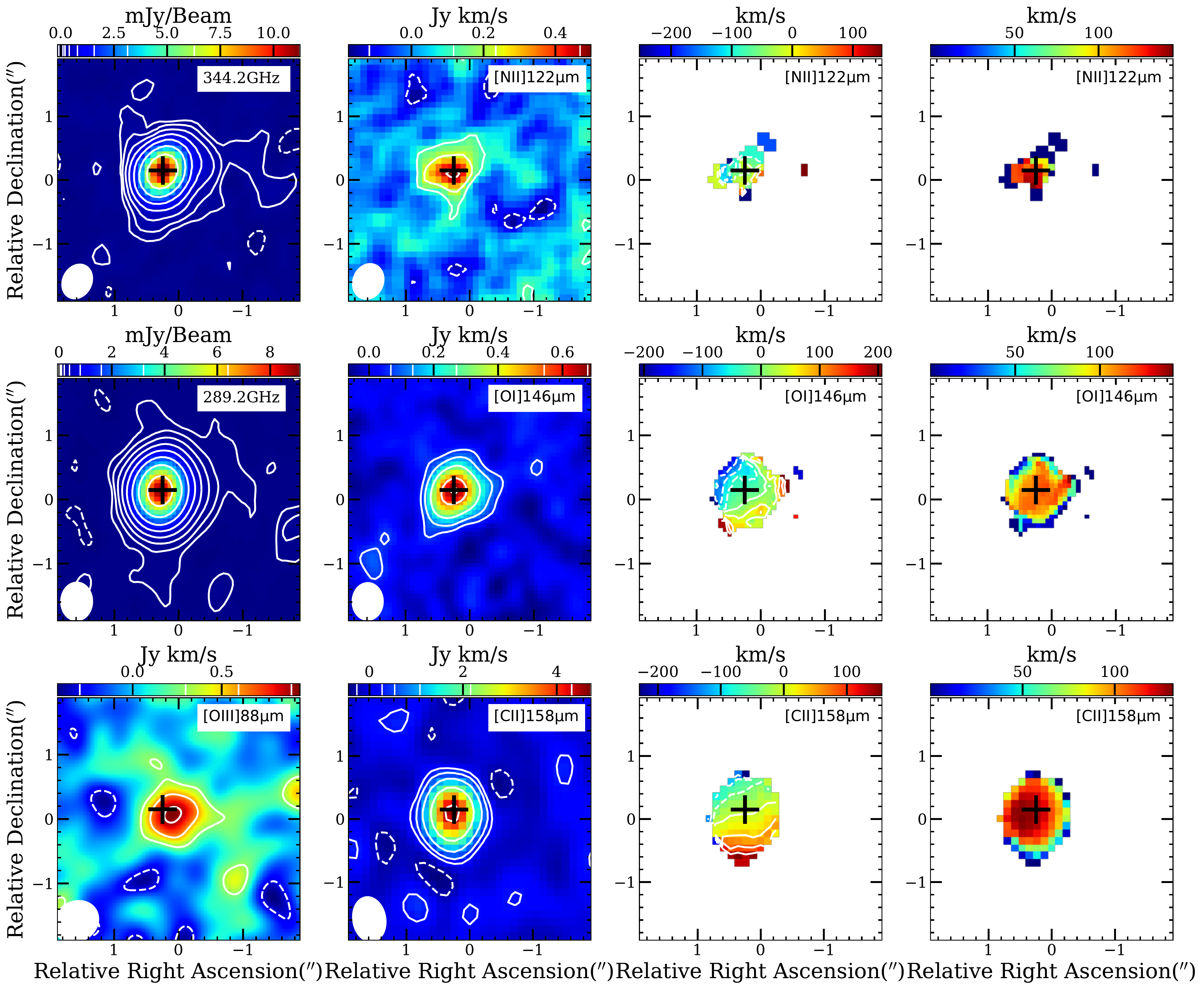}
 \caption{Continuum maps and  spectral line intensity, velocity, and velocity dispersion maps ({\bf from left to right}) of \nii{} and \oi{} (the first and second rows respectively, this work). The third row shows the \oiii{} intensity map \citep{hashimoto18}, and the \cii{} intensity map, velocity and velocity gradient maps  (\citealt{wang13}; {\bf from left to right}) for comparison.  
The black cross represents the HST position of the quasar (Shao et al. 2020 in preparation). The filled white ellipse on the lower left shows the FWHM of the beam.
The first column shows the continuum maps (the first and second row) and the intensity map of the \oiii{} line (the third row), and the white contours denote [ -2, 2, 4, 8, 16, 32, 64, 128 ] $\times \ \sigma\ (1 \sigma = 49\ \rm \mu Jy\ beam^{-1} )$ at 344.2 GHz, [ -2, 2, 4, 8, 16, 32, 64, 128, 256 ] $\times \ \sigma\ (1 \sigma = 25\ \rm \mu Jy\ beam^{-1} )$ at 289.2 GHz, and [ -2, 2, 4, 6 ] $\times \ \sigma\ (1 \sigma = 0.149\ \rm Jy \ beam^{-1}\cdot km \ s^{-1} )$ for the \oiii{} line, from top to bottom. 
The second column shows the intensity maps of the \nii{}, \oi{}, and \cii{} line, respectively. 
The white contours denote [ -2, 2, 4, 8 ] $\times \ \sigma\ (1 \sigma = 0.059\ \rm Jy \ beam^{-1}\cdot km \ s^{-1} )$ for the \nii{} line, [ -2, 2, 4, 8, 16, 20 ] $\times \ \sigma\ (1 \sigma = 0.034\ \rm Jy\  beam^{-1}\cdot km \ s^{-1} )$ for the \oi{} line, and  [ -2, 2, 4, 8, 16, 32 ] $\times \ \sigma\ (1 \sigma = 0.135\ \rm Jy\ beam^{-1} \cdot km \ s^{-1} )$ for the \cii{} line, from top to bottom.
The third and fourth column shows the velocity and velocity dispersion maps of the \nii{}, \oi{}, and \cii{} lines. The white contours on the velocity maps denote  [ -3, -2, -1, 0, 1 ] $\times \ \sigma\ (1 \sigma = 40\ \rm km \ s^{-1} )$ for the \nii{} line, [ -3, -2, -1, 0, 1 ,2, 3, 4 ] $\times \ \sigma\ (1 \sigma = 40\ \rm km \ s^{-1} )$ for the \oi{} line, and [ -3, -2, -1, 0, 1, 2, 3 ] $\times \ \sigma\ (1 \sigma = 40\ \rm km \ s^{-1} )$ for the \cii{} line, from top to bottom. 
}
\label{maps}
\end{figure*}

Figure \ref{maps}  shows the continuum, and the spectral line intensity, velocity, and velocity dispersion maps of the  \oi{} and  \nii{} lines. 
The maps of the \cii{} and \oiii{} lines are shown at the bottom for comparison. The peaks of all the continuum and spectral lines are consistent with the quasar's HST position (Shao et al. 2020 in preparation). We measured the source size by fitting a 2D Gaussian component to the intensity map. 
The \oi{} line is spatially resolved with a source size of ($0\farcs50$ $\pm$ $0\farcs07$) $\times$ ($0\farcs32$ $\pm$ $0\farcs09$) deconvolved from the beam. The \nii{} line is marginally resolved with a source size of ($0\farcs54$ $\pm$ $0\farcs19$) $\times$ ($0\farcs31$ $\pm$ $0\farcs16$). The measured source sizes from the \oi{} and \nii{} lines are comparable to those of the \cii{} line of ($0\farcs55$ $\pm$ $0\farcs05$) $\times$ ($0\farcs40$ $\pm$ $0\farcs07$) and the \oiii{} line of ($0\farcs44$ $\pm$ $0\farcs27$) $\times$ ($0\farcs38$ $\pm$ $0\farcs13$), and slightly larger than those observed in high$-J$ CO lines of ($0\farcs34$ $\pm$ $0\farcs11$) $\times$ ($0\farcs32$ $\pm$ $0\farcs16$) \citep{li20}.
The velocity field of the \oi{} line shows a velocity gradient from the northeast to the southwest that is similar to that of the \cii{} line. Considering the beam-smearing effect, the velocity gradient of \oi{} averaged in the outermost region of 32 \kmps{} is consistent with that of the \cii{} line of 61 \kmps{}. 
The underlying continuum emissions are all detected at very high S/Ns ($> 250$) \citep{shao19}.
The deconvolved source sizes for the continuum are all slightly smaller than those values determined for the emission lines (Tables \ref{tabline} and \ref{tabcont}).  

%More details are presented in Table \ref{tabline} and \ref{tabcont}.

\section{Discussion}

\subsection{Properties of \qso{}}
Our ALMA detection of the \oi{} and \nii{} emission lines together with previous observations of the \oiii{} and \cii{} lines in \qso{} allows us to constrain the atomic and ionized gas masses as well as the SFR.  We follow the method in \citet{ferkinhoff11} to estimate the ionized gas mass from the \nii{} and \oiii{} emission lines. 
Given the first and second nitrogen ionization energies of 14.5 and 29.6 eV, it is likely that in \hii{} regions, nitrogen is mostly singly ionized and traces much of the fully ionized hydrogen, and its abundance relative to H ($\chi (\rm N^{+})$) can be estimated from the nitrogen relative abundance (N/H) through $\chi (\rm N^{+})$ $\lesssim$ N/H.
For a local thermal equilibrium (LTE) level distribution, the line emission thus provides a lower limit on the ionized gas mass associated with the emission.
%Assuming a total nitrogen abundance $\chi (\rm N)$ $\lesssim$  N/H, 
The minimum ionized gas mass traced by singly ionized nitrogen is:
\begin{align}
M_{\rm min}(\rm H^{+}) = \frac{\textit{L}_{\rm [NII]_{122\mu m}} \textit{m}_{\rm H}}{\frac{\textit{g}_{2}}{\textit{Q}({\textit{T}_{\rm ex}})}   \textit{A}_{\rm 21} \textit{h} \nu_{21} \chi (N^{+})},
\label{eq1}
\end{align}
where $m_{\rm H}$ is the atomic hydrogen mass, $g_{2} = 5$ and $A_{21} = 7.5 \times 10^{-6}\ s^{-1}$ are the statistical weight and Einstein coefficient of the upper energy level, $Q({T_{\rm ex}}) = 1+ 5e^{-188/T_{\rm ex}} + 3e^{-70/T_{\rm ex}} = 9$ (at the high-temperature limit) is the partition function, $h$ is the Plank constant, and $\nu_{21}$ is the line frequency.  Adopting a solar nitrogen abundance $\chi (\rm N^{+})\lesssim N/H = 6.76 \times 10^{-5}$ from \citet{asplund09}, we derive a minimum ionized gas mass of $M^{min}_{ionized}= 6.2 \times 10^{8}$ $\rm M_{\sun}$.
%$\chi (\rm N^{+})= 9.3 \times 10^{-5}$ \citep{savage96}  $M^{min}_{atomic}= 4.5 \times 10^{8}$ \msun{}.
The ionized gas mass can also be derived from the \oiii{} line by replacing the \nii{} related terms with those of the \oiii{} line in Eq.\ref{eq1}. 
The parameters for \oiii{} are $g_{2} = 3$, $A_{21} = 2.7 \times 10^{-5}\ s^{-1}$, $Q({T_{\rm ex}}) = 1+ 5e^{-440/T_{\rm ex}} + 3e^{-163/T_{\rm ex}} = 9$ (at the high-temperature limit), and $\chi (\rm O^{++}) \textless O/H= 4.90 \times 10^{-4}$  (solar abundance; \citealt{asplund09}). 
Adopting an \oiii{} luminosity of $2.44 \times 10^{9}$ \lsun{} \citep{hashimoto18}, we estimate a minimum ionized gas mass of $M^{min}_{ionized}= 7.7 \times 10^{7}$ $\rm M_{\sun}$.
%$\chi (\rm O^{++})= 5.90 \times 10^{-4}$  \citep{savage96}  $M^{min}_{atomic}= 6.4 \times 10^{7}$ \msun{}.

In addition, we estimate the neutral gas mass from the \cii{} and \oi{} lines.
In the optically thin regime, the singly ionized carbon mass   (\citealt{weiss05}; \citealt{venemans17b}) is 
\begin{align}
\frac{M_{\rm C^{+}}}{M_{\sun}} = 2.92\times 10^{-4} \frac{Q(T_{\rm ex})}{4} e^{91.2/T_{\rm ex}} \frac{L_{\rm [CII]}^{'}}{\rm K\ km\ s^{-1}\ pc^{-2}},
\end{align}
where $Q(T_{\rm ex}) = 2 + 4e^{-91.2/T_{\rm ex}}$ is the \ion{C}{2} partition function,  and $T_{\rm ex}$ is the excitation temperature. 
Assuming LTE where $T_{\rm kin} \approx T_{\rm ex} $, and adopting a gas  temperature of $T_{\rm kin} = 228 \ \rm K$ derived from the CO spectral line energy distribution (SLED) \citep{li20}, we estimate the singly ionized carbon mass of $M_{\rm C^{+}}= 2.1\times 10^{7} \rm \ M_{\sun}$. 
With a $\chi(\rm C^{+})\lesssim C/H = 2.69 \times 10^{-4}$ (solar abundance; \citealt{asplund09}), we constrain the minimum atomic gas mass of $M^{min}_{neutral}= 7.8 \times 10^{10} \rm \ M_{\sun}$\footnote{If we adopted a temperature of $T_{\rm ex} = \rm 500,\ 100\ or\ 50 \  K$, the estimated gas mass would be $\rm 0.9,\ 1.3\ or \ 2.3 $ times of the value presented here.}.
Following a similar formula to \cii{}, the atomic oxygen mass can be calculated by substituting the parameters with those of the \oi{} line in Eq. 1 of \citet{weiss05}:
\begin{align}
\frac{M_{\rm O^{+}}}{M_{\sun}} = C m_{\rm OI} \frac{8\pi k {\nu_{0}}^{2}}{hc^{3}A_{01}}Q(T_{\rm ex})  e^{\frac{T_{\rm 0}}{T_{\rm ex}}}\frac{L_{\rm [OI]( ^{3}P_{0}- ^{3}P_{1})}^{'}}{\rm K\ km\ s^{-1}\ pc^{-2}}\\
= 6.19\times 10^{-5} Q(T_{\rm ex})e^{329/T_{\rm ex}} \frac{L_{\rm [OI]( ^{3}P_{0}- ^{3}P_{1})}^{'}}{\rm K\ km\ s^{-1}\ pc^{-2}},
\end{align}
where $Q(T_{\rm ex}) = 5 + 3e^{-228/T_{\rm ex}} + e^{-329/T_{\rm ex}}$ is the partition function of oxygen, $A_{01}$ is the Einstein coefficient, $T_{\rm 0}$ is the energy above the ground state, and $C$ is the conversion factor between $\rm pc^{2}$ and $\rm cm^{2}$. 
We derive an ionized oxygen mass of $M_{\rm O^{+}} = 8.1\times 10^{6} \rm \ M_{\sun}$ (for $T_{\rm ex}=228 \rm \ K$), which leads to $M^{min}_{neutral}= 1.7 \times 10^{10} \rm \ M_{\sun}$ adopting $\chi (\rm O^{+})\lesssim O/H = 4.90 \times 10^{-4}$ \footnote{If we adopted a temperature of $T = \rm 500,\ 100\ or\ 50 \  K$, the estimated gas mass would be $\rm 0.5,\ 5.3\ or \ 135.1$ times of the value presented here.}  (solar abundance; \citealt{asplund09}).
The estimated neutral gas masses from the \cii{} and \oi{} lines are comparable to the molecular gas mass derived from the CO$(2-1)$ line of $4.3\times 10^{10}$ \msun{} \citep{shao19}, and about an order of magnitude larger than the ionized gas mass derived from the \nii{} and \oiii{} lines.

The FIR FS lines are tracers of the SFR (e.g., \citealt{farrah13}; \citealt{delooze14}). We estimate the SFR based on the FS emission lines using Eq. 6 of \citet{farrah13}:
\begin{flalign}
& {\rm log10}(\frac{\dot{M_{\sun}}}{M_{\sun}\ \rm yr^{-1}}) =  (-7.02 \pm 1.25)&\notag\\
& \ \ \ \ \ \ \  \ \ \ \ \ \ \ \ \ \ \ \ \ \ +(1.07 \pm0.14){\rm log10}(L_{[\rm OIII]_{88\mu m}})&\notag\\
&=   (-7.30 \pm 0.87) + (1.19 \pm 0.11){\rm log10}(L_{[\rm NII]_{122\mu m}})&\notag\\
&=   (-10.04 \pm 1.34) + (1.55 \pm 0.17){\rm log10}(L_{[\rm OI]_{146\mu m}})&\notag\\
&=   (-6.24 \pm 1.72) + (0.95 \pm 0.19){\rm log10}(L_{[\rm CII]_{158\mu m}})&\notag
\end{flalign}
The SFRs estimated from the \cii{}, \oi{}, \nii{}, and \oiii{} lines are 1524, 11917, 2208, and 1058 $\rm M_{\sun}\ yr^{-1}$, respectively. 
%The \oi{} line suggests the highest estimated SFR, but is within the statistical uncertainties compared to the other FS lines\footnote{The statistical uncertainties are 2.6, 2.0, 1.3, and 1.8 dex for the \cii{}, \oi{}, \nii{}, and \oiii{} lines estimated from the relation \citep{farrah13}.}. The high SFR derived from the \oi{} line maybe a result of (i) extra gas heating by the AGN in addition to the UV heating from young massive stars, evident from the higher $L_{\rm [OI]_{146\mu m}/} L_{\rm FIR}$ ratio in \qso{} compared to local (U)LIRGs and AGNs, that is not found for other FS lines (see discussion in Section \ref{firdeficit}), and (2) the large uncertainty of extrapolating the relation between the \oi{}  luminosity and the SFR in \citet{farrah13} to a luminosity of  $\sim 10^{9}$ \lsun{} (we note that in \citealt{farrah13}, the relation is based on sources with \oi{} luminosity less than  $10^{9}$ \lsun{}). 
We note that the \oi{} line suggests an extremely high SFR. The error bars of the parameters in the above relations are large, which result in large uncertainties in the SFR estimation. \footnote{The statistical uncertainties are 2.6, 2.0, 1.3, and 1.8 dex for the \cii{}, \oi{}, \nii{}, and \oiii{} lines estimated from the relation \citep{farrah13}.} Moreover, the \oi{} line flux may not give a reliable estimate of the SFR, as (1) the high $L_{\rm [OI]_{146\mu m}/} L_{\rm FIR}$ ratio (see discussions in Section \ref{firdeficit}) suggests extra gas heating by the AGN in addition to the UV photons from young massive stars, and (2) the relation in \citet{farrah13} is based on samples with \oi{} luminosity less than  $10^{9}$ \lsun{}, which may not be valid when extrapolated to higher \oi{} luminosity for SFR calculation.
The SFR derived from the infrared dust ($10^{-11} \times L_{\rm IR}$) ranges from 1300 to 3600 $\rm M_{\sun}\ yr^{-1}$, depending on the different model parameters used in the dust continuum SED fitting \citep{shao19}. The SFRs traced by both the dust continuum and the \cii{}, \nii{} and \oiii{} FIR FS lines are consistent, while  the \oi{} line suggests a higher SFR, although all the derived values are consistent within statistical uncertainties. The derived gas masses and SFRs are listed in Table \ref{tabline}.
\subsection{Line-to-FIR luminosity ratios}\label{firdeficit}
\begin{figure*}
\centering 
\includegraphics[width=1.\textwidth]{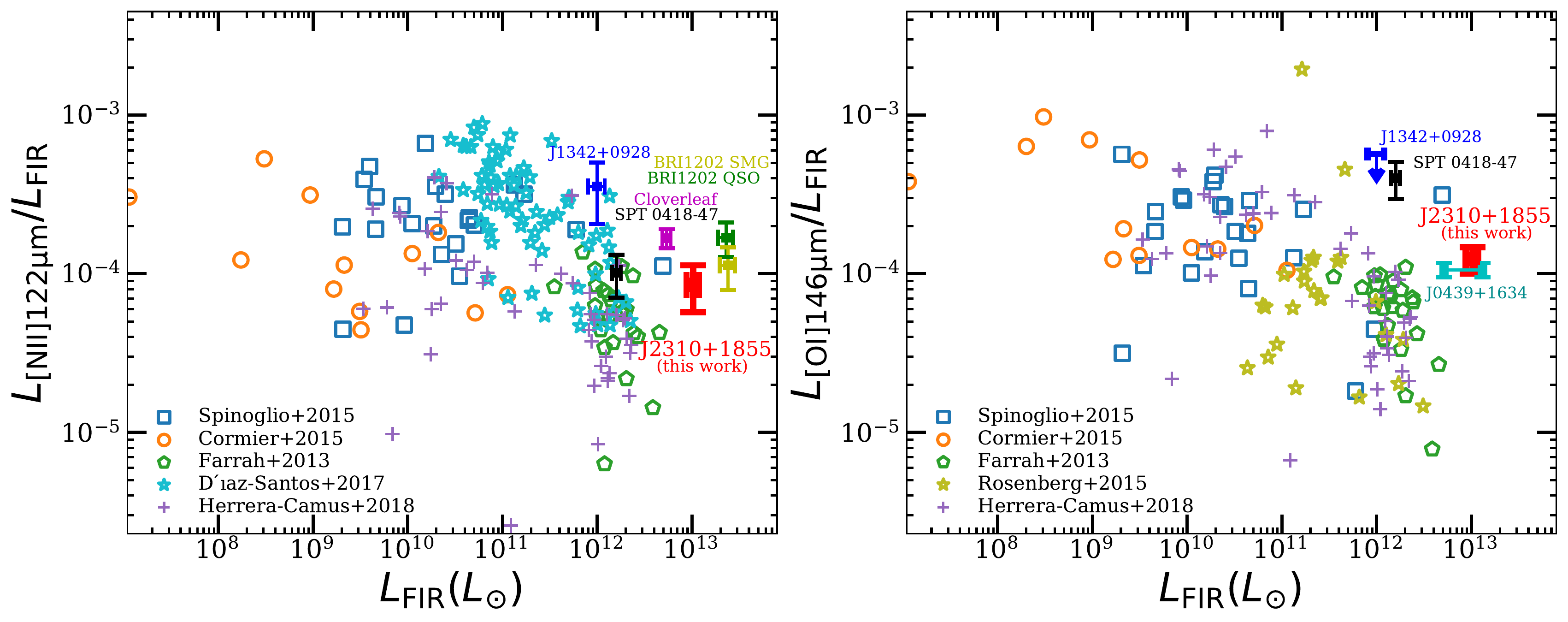}
\includegraphics[width=1.\textwidth]{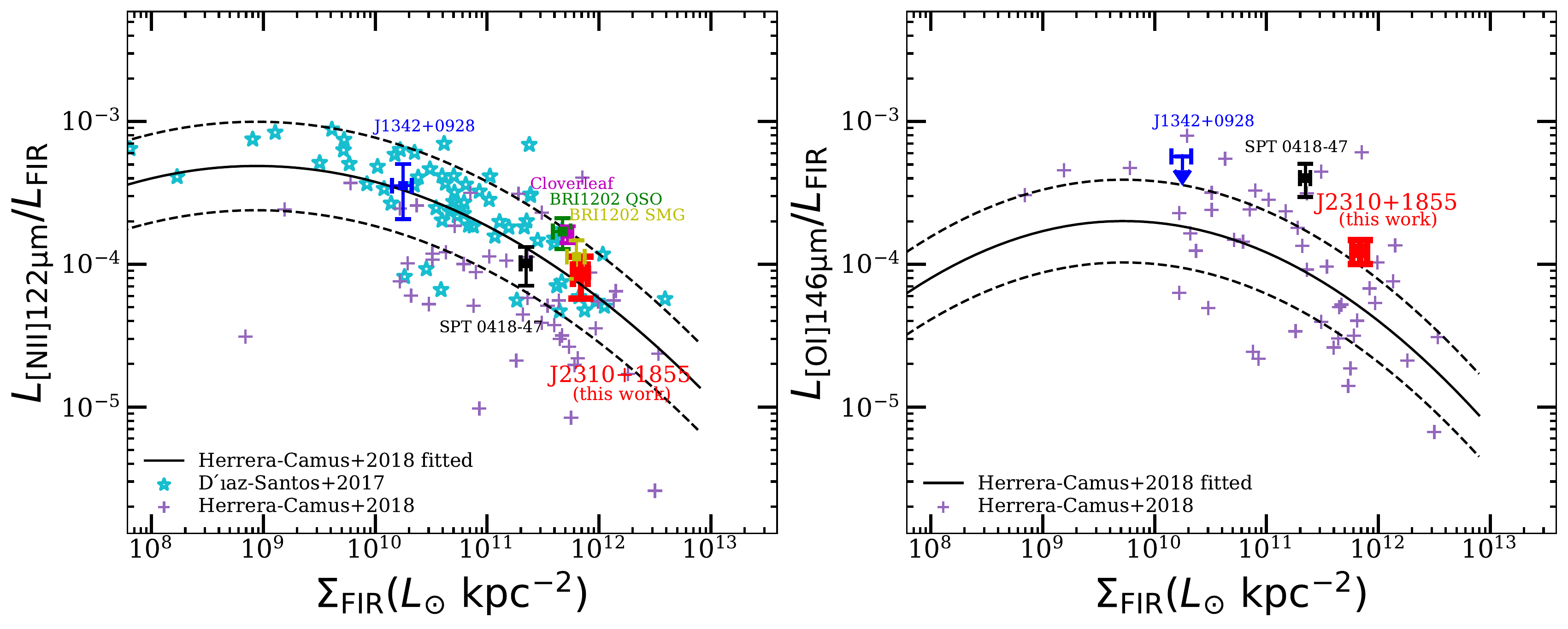}
 \caption{Top:  \nii{} (left) and \oi{} (right) to FIR luminosity (42.5 $-$ 122.5 $\mu m$) ratio as a  function of FIR luminosity for local galaxy samples and high-$z$ sources. 
 The local galaxy samples include: 26 Seyfert galaxies from \citet{spinoglio15}; 40 low-metallicity dwarf galaxies from \citet{cormier15}; 25 (U)LIRGs (consisting of 6 galaxies and 19 AGNs) from \citet{farrah13}; 240 LIRGs from \citet{diaz17}; 29 (U)LIRGs (consisting of 16 AGNs and 13 starburst galaxies) from \citet{rosenberg15}; a composite sample of AGNs, star-forming galaxies and LIRGs with a total number of 52 galaxies from \citet{herrera18a}. We note that all the plotted data points in the samples are those with the \nii{} or the \oi{} line detected. 
The high-redshift sources are $z=7.54$ quasar J1342+0928 \citep{novak19},  the quasar J0439+1634 at $z=6.52$ \citep{yang19}, $z=4.22$ lensed dusty star-forming galaxy SPT 0418-47 \citep{debreuck19}, the $z=2.56$ quasar the Cloverleaf \citep{ferkinhoff15}, and the $z=4.69$ quasar BRI1202 QSO and its companion galaxy BRI1202 SMG \citep{lee19}.
The quasar \qso{} (this work) is shown as a red square with calibration uncertainties of 10$\%$ for the \nii{} and \oi{} lines included, respectively. The connected cyan squares demonstrate the range of intrinsic FIR luminosity of the quasar J0439+1634 (corrected for lensing with $\mu = 2.6- 6.6$, \citealt{yang19}).
Bottom:   \nii{} (left) and \oi{} (right) to FIR luminosity ratio as a function of FIR surface brightness for  local galaxy samples and high-$z$ sources. The black solid lines are the best fit to the AGNs, star-forming galaxies, and LIRGs in \citet{herrera18a}, and the black dashed lines demonstrate the uncertainties. The source sizes used to calculate the FIR surface brightness for the high-$z$ sources are the continuum size at 344.2 GHz for \qso{} (this work), the radius reconstructed from the lensing model for SPT 0418-47 \citep{gullberg15},  the \cii{} continuum size for J1342+0928  \citep{banados19}, and the the \cii{} continuum size for the Cloverleaf quasar \citep{uzgil16}. We note that the FIR luminosities in  \citet{cormier15} and \citet{farrah13} are converted from the infrared luminosity (8 $-$ 1000 $\mu m$) assuming an IR/FIR ratio of 1.75.
}
\label{line_to_ir}
\end{figure*}

We compare the \nii{} and \oi{} emission in  \qso{} with values found for local and high-$z$ systems by studying the line-to-FIR ratio as a function of FIR luminosity ($L_{\rm FIR}$), and as a function of FIR surface brightness ($\Sigma_{\rm FIR}$). 
The comparison samples include samples of local dwarf galaxies, (U)LIRGs, and AGNs (\citealt{farrah13}; \citealt{cormier15}; \citealt{spinoglio15}; \citealt{rosenberg15}; \citealt{diaz17}; \citealt{herrera18a}). We also include high-$z$ objects with available \oiii{} and \nii{} detections for comparison: the $z=7.54$ quasar J1342+0928 \citep{novak19}, the quasar J0439+1634 at $z=6.52$ \citep{yang19},
  the Cloverleaf, a lensed  $z=2.56$ quasar \citep{ferkinhoff15}, the $z=4.22$ lensed dusty star-forming galaxy SPT 0418-47 \citep{debreuck19}, and the $z=4.69$ quasar BRI1202 QSO and its companion submillimeter galaxy (SMG) BRI1202 SMG \citep{lee19}.
Detailed descriptions about the comparison samples are provided in the caption of  Figure  \ref{line_to_ir}.

We show the results in Figure  \ref{line_to_ir}.
For the \nii{} line, all the local (U)LIRGs and AGNs show decreasing $L_{\rm [NII]_{122\mu m}/} L_{\rm FIR}$ ratios with increasing $L_{\rm FIR}$ for $L_{\rm FIR}\gtrsim 10^{11}$ \lsun{}. 
%The samples with AGNs exhibit more of a ``deficit" compared to the (U)LIRGs, which maybe attributed to the contribution of the AGN to the FIR luminosity.
We adopt the \qso{}  FIR luminosity of $L_{\rm FIR} = (1.03 \pm 0.22) \times 10^{13}$ \lsun{}  from \citet{shao19}.
The quasar \qso{}, together with other high-z objects (e.g., the Cloverleaf,  BRI1202 SMG and  BRI1202 QSO) show a slightly higher $L_{\rm [NII]_{122\mu m}/} L_{\rm FIR}$ ratio than the local (U)LIRG and AGN samples for $L_{\rm FIR} \sim 10^{13} L_{\sun}$,  while the high-z objects with $L_{\rm FIR} \sim 10^{12} L_{\sun}$ (e.g., J1342+0928 and SPT 0418-47), follow the trend of the local (U)LIRG and AGN samples.
%The other two high-$z$ objects J1342+0928 and SPT 0418-47 with  $L_{\rm FIR}$ of $ \sim 10^{12} L_{\sun}$ have $L_{\rm [NII]_{122\mu m}/} L_{\rm FIR}$ ratios within the range of the local (U)LIRGs and AGNs at the same FIR luminosity level.
Recent studies find that the ``line deficit" is more closely correlated with $\Sigma_{\rm FIR}$ than $L_{\rm FIR}$ in samples of local (U)LIRGs and AGNs (e.g., \citealt{lutz16}; \citealt{diaz17}; \citealt{herrera18a}; \citealt{herrera18b}).
For all the local comparison samples,  FIR source sizes are only available in \citet{diaz17} and \citet{herrera18a}. We compare the  line-to-FIR ratio of the quasar with the above two samples of local (U)LIRGs and AGNs as a function of $\Sigma_{\rm FIR}$, and the results are shown in Figure  \ref{line_to_ir}.
%We select from comparison samples the far infrared characteristics (source size) that have been provided, and the result is shown in Fig  \ref{line_to_ir}.
For the high-$z$ objects, we calculate $\Sigma_{\rm FIR}$ following the definition of \citet{herrera18a} and \citet{diaz17}.
All of the high-z objects including \qso{}, lie within the range of (U)LIRGs and AGNs.
This suggest that  the $L_{\rm [NII]_{122\mu m}/} L_{\rm FIR}$ ratio measured from the nuclear region ($\lesssim 3$ kpc) of \qso{} is comparable to that of local (U)LIRGs, AGNs and other high-$z$ systems with a similar FIR surface brightness.
%\qso{}, together with J1342+0928 and SPT 0418-47, follows the trend of local (U)LIRGs and AGNs. The significantly higher $L_{\rm [NII]_{122\mu m}/} L_{\rm FIR}$ ratio found in the case of the Cloverleaf at its $\Sigma_{\rm FIR}$ compared to local systems, is likely a result of XDR contribution to the \nii{} line, considering the quasar is strongly lensed and the emission is coming from the very center of the galaxy ($r \textless 650$ pc; \citealt{ferkinhoff15}; \citealt{uzgil16}). 

As for the \oi{} line, all of the local (U)LIRGs and AGNs  reveal decreasing $L_{\rm [OI]_{146\mu m}/} L_{\rm FIR}$ ratios with increasing $L_{\rm FIR}$ for $L_{\rm FIR}\gtrsim 10^{11}$ \lsun{} (Figure  \ref{line_to_ir}).
\qso{}, together with J1342+0928 and SPT 0418-47, indicate higher $L_{\rm [OI]_{146\mu m}/} L_{\rm FIR}$ ratios compared to local samples at similar $L_{\rm FIR}$.
As for J0439+1634, the uncertainty in the magnification factor makes it hard to conclude if values are consistent with the local relation (for $\mu_{\rm max} = 6.6$) or if it has a comparable $L_{\rm [OI]_{146\mu m}/} L_{\rm FIR}$ ratio to \qso{} at similar $ L_{\rm FIR}$ (for $\mu_{\rm min} = 2.6$, \citealt{yang19}).
As for the case of $\Sigma_{\rm FIR}$, similar to other high-$z$ quasars and galaxies, the \oi{} line flux measured in the central $\approx 3$ kpc region of \qso{} exceeds the range of local systems and exhibits over two times the values in local (U)LIRGs and AGNs measured from the galaxy scale, suggesting no sign of \oi{} deficit. The \oi{} emission traces the warm and dense neutral gas (with temperature $\rm T \sim \Delta E/K = 329 \ K$ and critical density $\rm n_{cr} = 9.4 \times 10^{4} \ cm^{-3}$). The luminous \oi{} detection of \qso{} suggests warm and dense gas in the nuclear region, which is possibly heated by the photons contributed by the AGN in addition to the UV photons from young massive stars. This is in agreement with spatially resolved observations of local AGNs, with higher \oi{} fluxes detected in the central regions (which suggest warm and dense gas) compared to the mean values found in the outer star-forming disk \citep{herrera18b}.
In addition, we compare the \cii{} and \oiii{} lines of \qso{} with the $L_{\rm line}/ L_{\rm FIR}$ versus $\Sigma_{\rm FIR}$ relation observed in local galaxies, (U)LIRGs and AGNs \citep{herrera18a}. 
The results suggest that the \cii{}, \nii{}, and \oiii{} lines exhibit a similar deficit as in local systems (\citet{wang13}; \citet{hashimoto18}). The \oi{} line is over two times brighter than in local (U)LIRGs and AGNs at its $\Sigma_{\rm FIR}$, suggesting that the absence of an \oi{} deficit could be caused by a warm and dense neutral gas content, which is possibly heated by the photons from the AGN in addition to the UV photons from young massive stars. 
\subsection{ \cii{} from the ionized gas} \label{hiicii}
The \cii{} line traces both the neutral and the ionized ISM. The contribution to the \cii{} emission from the HII regions can be constrained if tracers purely from ionized gas phase are observed. The \nii{} and \niis{} lines, as examples, are widely used to probe the \cii{} emission from the ionized gas phase in the local universe to the highest redshift ($z\gtrsim 6$; e.g., \citealt{oberst06, oberst11}; \citealt{herrera16,herrera18a}; \citealt{diaz17}; \citealt{novak19}). 
The  \nii{} and \niis{} lines are from the same ionization state of nitrogen, thus the ratio between these two lines has a negligible dependence on the radiation field and metallicity. The difference in the critical densities between the \nii{} (310 \pcubcm{}) and \niis{} (48 \pcubcm{}) lines enables the \nii{}/\niis{} ratio to trace  ionized gas density.
Once the gas density is determined, the  \cii{}/ \niis{} ratio and the fraction of \cii{} emission from the ionized gas phase will be constrained by comparing the observational data with radiative transfer models.
%Using the \nii{} and \niis{} lines, we constrain the \cii{} from ionized gas following the steps below.
%Firstly,  gas density is estimated by the \nii{}/\niis{} ratio; Secondly, the \cii{}/ \niis{} ratio is predicted with the derived gas density when compared to theroetical models; Finally, the \cii{} from the ionized gas is constrained. 
%This method is valid even only the \niis{} line is observed. This is because that the \cii{} from ionized gas has a similar critical density (50 \pcubcm{}) as that of the \niis{} line, which leads to a mild dependence of the  \cii{}/ \niis{} ratio on gas density. 

With only the \nii{} line observed in \qso{}, we estimate its \niis{} luminosity from (1) the relation between the $L_{\rm [NII]_{205 \mu m}}/L_{\rm FIR}$ ratio and the FIR color ($f_{70}/f_{160}$) in local (U)LIRGs (Eq. 1 from \citealt{zhao16}), and (2) the relation between the $L_{\rm [NII]_{205 \mu m}}/L_{\rm [CII]_{158 \mu m}}$ ratio and the  FIR color ($f_{60}/f_{100}$) in local (U)LIRGs (Eq. 5 from \citealt{lu15}).
Adopting a dust temperature of 40 K and an emissivity index $\beta=1.6$ from \cite{shao19}, we estimate an  $f_{70}/f_{160}$ ratio of 2.23 and an $f_{60}/f_{100}$ ratio of 0.92. 
The two methods give comparable estimated \niis{} luminosities, with $L_{\rm [NII]_{205 \mu m}} = 4.2 \times 10^{8} \ \rm L_{\sun}$ derived from the infrared luminosity (method 1),  and $L_{\rm [NII]_{205 \mu m}} = 4.6 \times 10^{8} \ \rm L_{\sun}$ based on the \cii{} luminosity \footnote{The estimated \nii{}/\cii{} ratio in \qso{} of 1/18 is consistent with values found for other high-z quasars, e.g., BRI 1335--0417 at z = 4.41 and BRI 1202--0725 at z = 4.70 \citep{lu17,lu18}, as well as other high-z galaxies, e.g.,  ID 141 at z=4.24 \citep{cheng20}.} (method 2).  
Figure \ref{diagnostics} (a) shows the \cii{}/ \niis{} and \nii{}/ \niis{} ratios in the \hii{} region predicted by the radiative transfer code Cloudy \citep{ferland17}.
Comparing the estimated \nii{}/ \niis{} ratio \footnote{The \niis{} luminosity is estimated from the relation of local (U)LIRGs (PDRs). In  AGNs, the XDRs may contribute significantly to the \nii{} emission, which should be subtracted when using the  \nii{}/ \niis{} ratio as a density probe. In the case of \qso{}, the observed \nii{}/IR ratio is comparable to that found for local (U)LIRGs and AGNs, which disfavors a significant contribution of the \nii{} emission from XDRs. Accordingly the estimated \nii{}/\niis{} ratio here traces the ionized gas density.} with the model, we estimate that the ionized gas density is 45 $\rm cm^{-3}$ in \qso{}. 
We emphasis that if we adopt a hotter dust temperature than those suggested by \citet{wangf19}, the resulting estimated \niis{} luminosity will be lower.
Accordingly, we consider the derived ionized gas density as a lower limit.
The ionized gas density $n_{e} \gtrsim $45 $\rm cm^{-3}$ of \qso{} is consistent with that found in local (U)LIRGs and AGNs (10 $\sim$ 200 \pcubcm{} from  \citealt{herrera18a};  a median of 46 \pcubcm{} from \citealt{diaz17}), but lower than that in the $z=7.5$ quasar J1342+0928 of  \nele{} $\geq$ 180 \pcubcm{}.
With the estimated ionized gas density of $n_{e} \gtrsim $45 $\rm cm^{-3}$, we derive an ionized gas contribution of $\lesssim$17$\%$ to the \cii{} line in \qso. This is in agreement with the values derived from local (U)LIRGs and AGNs (\citealt{diaz17}; \citealt{herrera18a}), and the z= 7.54 quasar J1342+0928.

\subsection{Gas-phase Metallicity} \label{metallicity}
The \oiii{} and \nii{} lines have similar critical densities but different ionization potentials ($n_{cr} = 510\ \rm cm^{-3}$, E = 35.1 eV for \oiii{}, and $n_{cr} = 310\ \rm cm^{-3}$, E = 14.5 eV for \nii{}). Thus, the flux ratio between these two lines could be used to trace the ionization parameters and metallicity.
Utilizing the radiative transfer model Cloudy, \citet{pereira17} and \citet{rigopoulou18} find that  once the ionization parameter is determined, the ratio between the \oiii{} and \nii{} line can be used as a gas-phase metallicity indicator in star-forming galaxies. 
The detection of both lines in \qso{} allows us therefore to constrain the ISM metallicity.
However, several issues are noted before applying this method to quasars: (1) these models consider the galaxy stellar emission as the only radiation field. \citet{pereira17} predict that in the presence of an AGN, the \oiii{} emission is barely affected but the \nii{} line may have a significant contribution from XDRs (up to $\sim 90\%$) depending on the AGN power-law index; (2) the \oiii{}/\nii{} ratio also has strong dependence on the ionization parameter ($U$) of the radiation field. \citet{rigopoulou18} proposed using the ratio of underlying continuum  luminosity measured at the wavelengths of the \oiii{} and \nii{} lines ($\rm 88\mu m/122\mu m$) for the ionization parameter estimation. However, it should be noted that the observed continuum $\rm 88\mu m/122\mu m$ ratio does not reliably probe the ionization parameter of the galaxy stellar radiation field for systems with powerful AGNs, and (3) the strengths of the \oiii{} and \nii{} lines can be affected by the dust opacity effect. The local ULIRG  Arp 220, as an example, was detected in absorption for the  \oione{} and \nii{} FS lines (\citealt{rangwala11}; \citealt{gonz12}).  Optically thick dust for $\lambda_{rest} \lesssim 200 \mu m$ has also been observed in high-z SMGs \citep{riechers13}. It is also possible that the dust  in \qso{} is optically thick for $\lambda_{rest} \lesssim 200 \mu m$. However, current dust continuum observations are insufficient to constrain the dust opacity, as well as to correct for the extinction effect for the FS lines in \qso{}.

Following the method in \citet{rigopoulou18}, we first estimate the ionization parameter of the galaxy stellar radiation field from the continuum $\rm 88\mu m/122\mu m$ ratio. To exclude continuum emission contributed by the quasar, we use the decomposed galaxy dust emission with a dust temperature $T_{\rm dust} = 40\ \rm K$ and emissivity index $\beta = 1.6$ \citep{shao19} to estimate the continuum $\rm 88\mu m/122\mu m$  ratio. % instead of using the observed values.
%nstead of using the observed continuum $\rm 88\mu m/122\mu m$ ratio, we adopt the decomposed dust emission from the galaxy with a dust temperature $T_{\rm dust} = 40\ \rm K$ and emissivity index $\beta = 1.6$ \citep{shao19}. 
This leads to a continuum $\rm 88\mu m/122\mu m$ ratio of 1.4 and ionization parameter $\rm log(U) =  -2.6 \sim -2.1$ for typical \ion{H}{2} region density of $n_{e} = \rm 10^{1-3} cm^{-3}$ (Figure 4, \citealt{rigopoulou18}). 
From the grid presented in Table A1 of  \citet{pereira17}, the observed \oiii{}/\nii{} ratio of 2.8 in \qso{} corresponds to a metallicity of  $Z/Z_{\sun}$=1.5$-$2.1 \footnote{We note that if the AGN contributes significantly to the observed \nii{} flux, the estimated metallicity will be lower. For \qso{}, the comparable $L_{\rm [NII]_{122\mu m}/} L_{\rm FIR}$ ratio with (U)LIRGs suggests no clear sign of significant XDR contribution to the \nii{} line. } relative to solar metallicity for the estimated $\rm log(U) =  -2.6 \sim -2.1$. 
The metallicity  is consistent with that of other high-redshift quasars and galaxies ($Z/Z_{\sun}$=0.7$-$2.0 for J1342+0928, \citealt{novak19}; $Z/Z_{\sun}$=0.3$-$1.3 for SPT 0418-47, \citealt{debreuck19}).

\subsection{Diagnostics of the neutral and ionized ISM}
Combinations of different FIR FS emission lines provide a wealth of information about the ISM physical and chemical conditions, e.g., temperature, density, ionization parameter, metallicity, and volume filling factor ratio between different gas phases.
\qso{} has been detected in the \cii{}, \oi{}, \nii{}, and \oiii{} FS lines, making it possible to compare its physical conditions of the neutral and ionized gas with that of local and high-redshift systems.
The comparison samples are local dwarf galaxies, (U)LIRGs, and AGNs (\citealt{farrah13}; \citealt{cormier15}; \citealt{spinoglio15}; \citealt{fernandez16}; \citealt{herrera18a}), and the high-$z$ systems with available detections, including the quasars J1342+0928 and J0439+1634, as well as the lensed galaxy  SPT 0418-47.

\begin{figure*}
\gridline{\fig{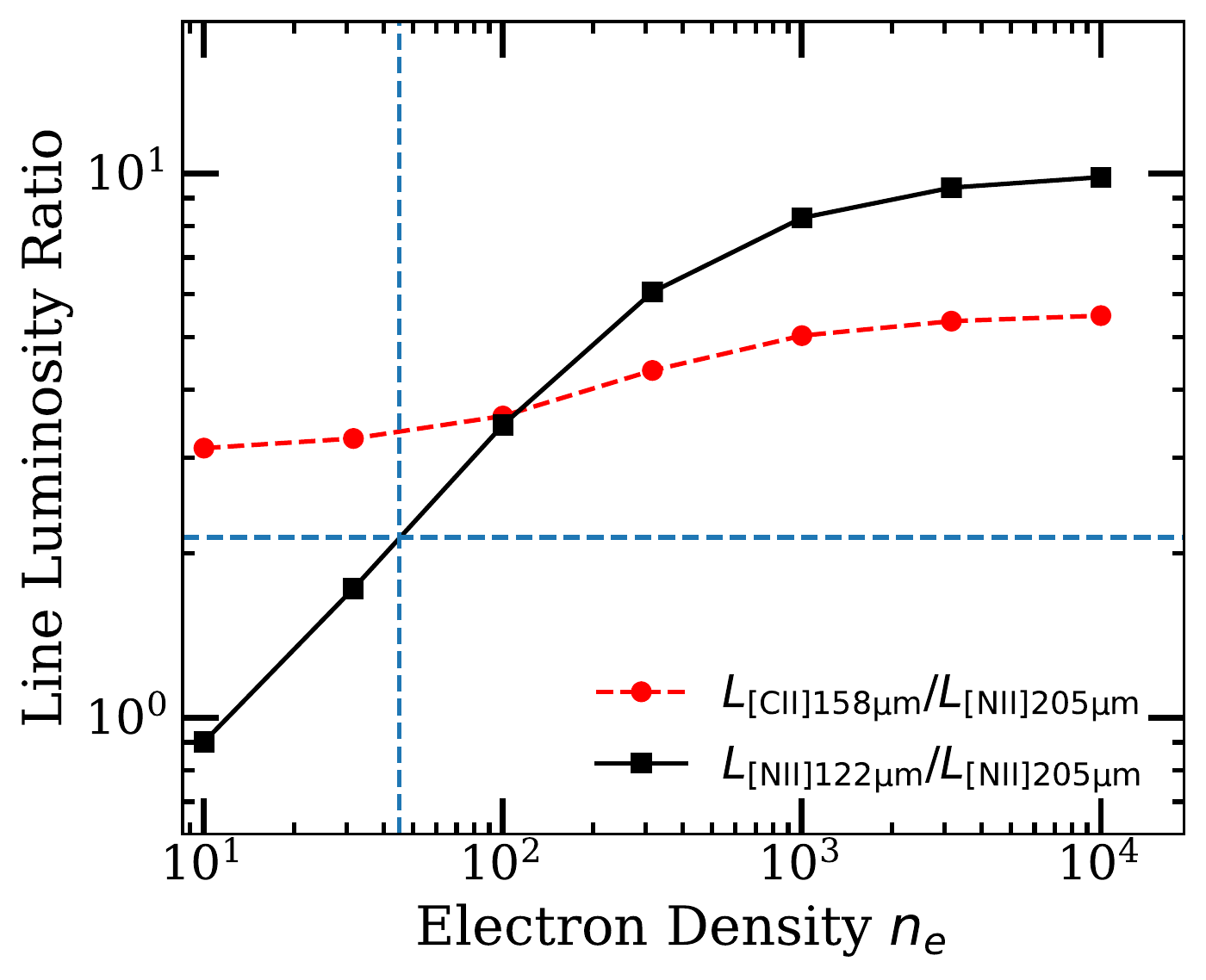}{0.46\textwidth}{(a)}
          \fig{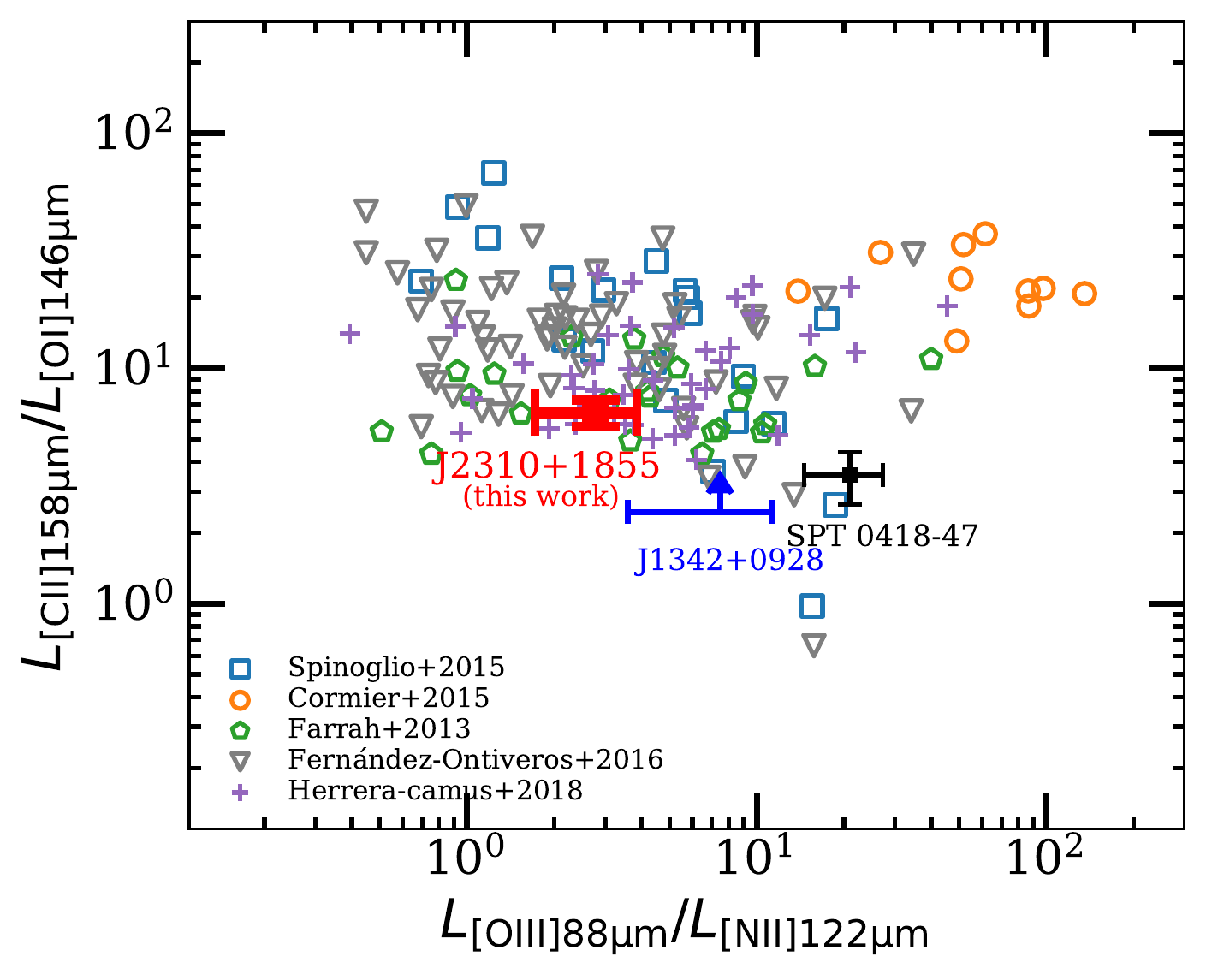}{0.46\textwidth}{(b)}
          }
\gridline{\fig{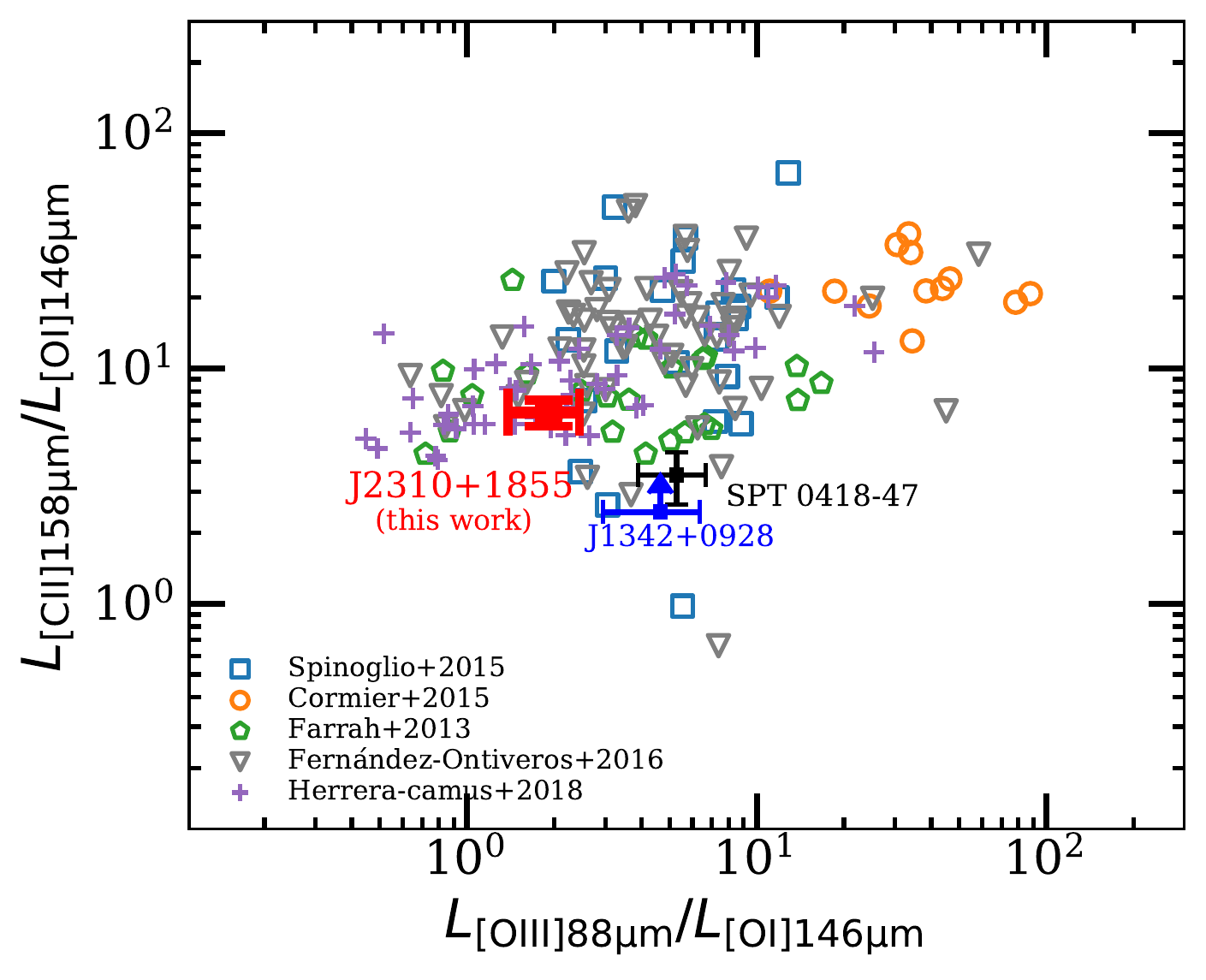}{0.46\textwidth}{(c)}
          \fig{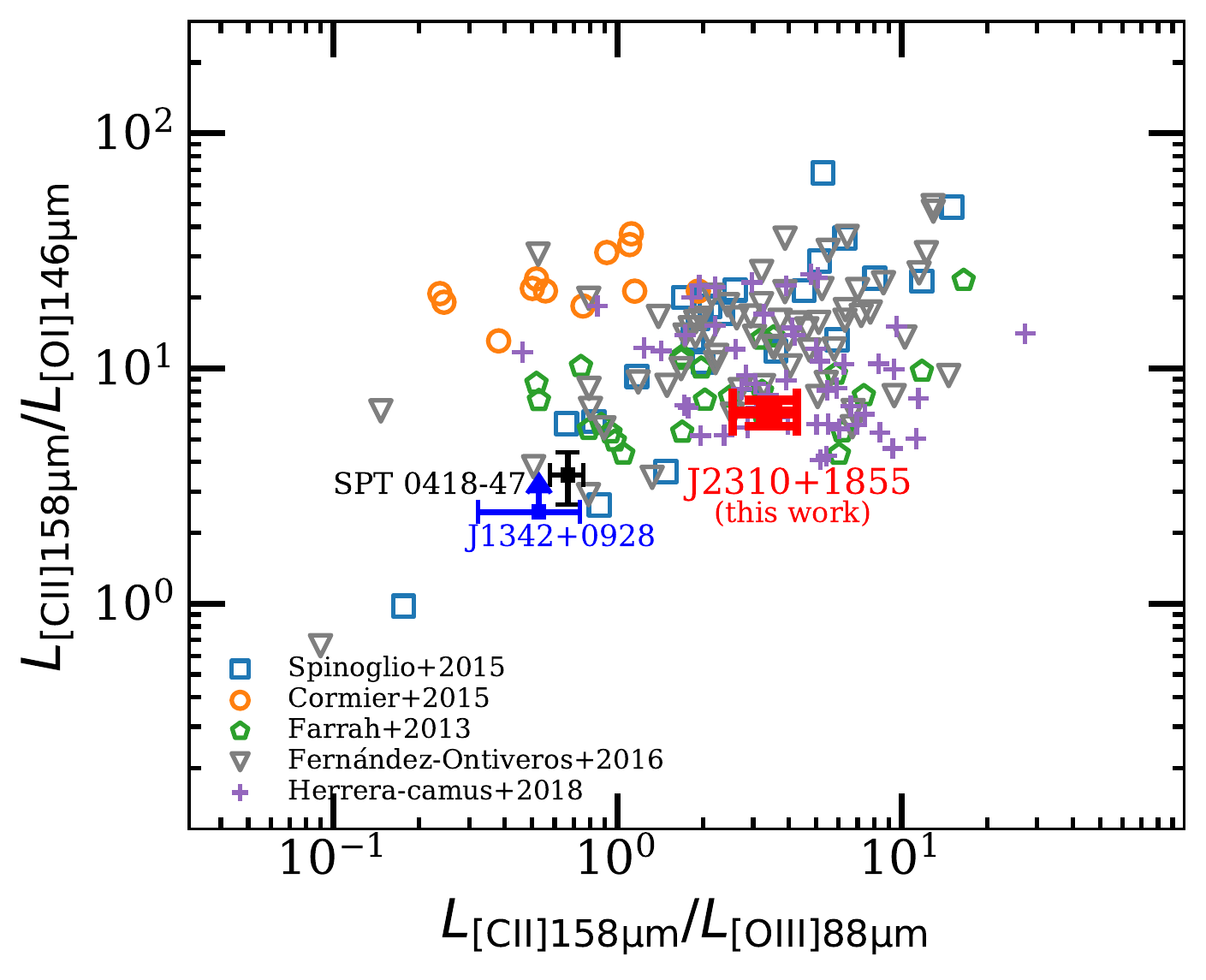}{0.46\textwidth}{(d)}
          }
\caption{(a). Cloudy \citep{ferland17} prediction of the \cii{} /\niis{} (red dashed line)\ and \nii{} /\niis{} (black solid line)\ intensity ratio in the \hii{} region for a typical solar metallicity stellar SED with an ionization parameter of $\rm U=-2$. The estimated \nii{} /\niis{} \ ratio suggests an electron density of \nele{} =45 \pcubcm{} (vertical dark blue dashed line). For the estimated electron density of \nele{} =45 \pcubcm{} , the predicted fraction of \cii{} from the HII region is 17$\%$. (b) $-$ (d) FS line ratio diagrams.
The local samples include 26 Seyfert galaxies from \citet{spinoglio15}; 40 low-metallicity dwarf galaxies from \citet{cormier15}; 25 ULIRGs (consisting of 6 galaxies and 19 AGNs) from \citet{farrah13};   a composite sample of 170 AGNs, 20 starburst galaxies, and 43 dwarf galaxies from \citet{fernandez16}; 29 (U)LIRGs, consisting of 16 AGNs and 13 starburst galaxies from \citet{rosenberg15}; and a composite of AGNs, star-forming galaxies and LIRGs with a total number of 52 galaxies from \citet{herrera18a}. We note that all the plotted data points in the samples are those with the fine-structure lines (e.g., \nii{},  \oi{}, \oiii{}, and \nii{}) detected.
The high-redshift sources are $z=7.54$ quasar J1342+0928 \citep{novak19} and $z=4.22$ lensed dusty star-forming galaxy SPT 0418-47 \citep{debreuck19}. The quasar \qso{} (this work) is shown as a red square with calibration uncertainties of 10$\%$ for the \nii{} and \oi{} lines included, respectively.
}
\label{diagnostics}
\end{figure*}

The majority of the \cii{} emission is originating in the neutral gas phase. The \oi{} line comes purely from the neutral gas phase and requires higher temperature and higher density to be excited compared to the \cii{} line ($\rm T \sim \Delta E/k = 329 \ K$, $\rm n_{cr} = 9.4 \times 10^{4} \ cm^{-3}$ for \oi{} and $\rm T \sim \Delta E/k = 91 \ K$, $\rm n_{cr} = 2.8 \times 10^{3} \ cm^{-3}$ for \cii{}). 
Thus the \cii{}/\oi{} ratio depends on both temperature and atomic gas density. The  \nii{} and \oiii{} lines trace the ionized gas phase, and the \oiii{}/\nii{} ratio is dependent on metallicity and ionization parameter (Section \ref{metallicity}). 
%Having similar critical densities but different ionization potential ($\sim 500\ \rm cm^{-3}$, 35.1 eV for \oiii{}, and $\sim 300\ \rm cm^{-3}$, 14.5 eV for \nii{}), the \oiii{}/\nii{} ratio traces the ionization parameter and also has strong dependence on metallicity (\citealt{nagao11}; \citealt{pereira17}; \citealt{{rigopoulou18}}).
Figure \ref{diagnostics}(b) shows the $L_{\rm [CII]_{158\mu m}}/ L_{\rm [OI]_{146\mu m}}$ vs $L_{\rm [OIII]_{88\mu m}}/ L_{\rm [NII]_{122\mu m}}$ diagnostic diagram.
The  $L_{\rm [CII]_{158\mu m}}/ L_{\rm [OI]_{146\mu m}}$ ratio is found to be lower in the high-temperature and high-density regime (\citealt{diaz17}; \citealt{herrera18a}). 
The AGN samples display a wide spread in the $L_{\rm [CII]_{158\mu m}}/ L_{\rm [OI]_{146\mu m}}$ ratio, while the (U)LIRGs and dwarf galaxies show less scatter with a median value of  $L_{\rm [CII]_{158\mu m}}/ L_{\rm [OI]_{146\mu m}}\sim 10$ for the (U)LIRGs and $L_{\rm [CII]_{158\mu m}}/ L_{\rm [OI]_{146\mu m}}\sim 20$ for the dwarf galaxies. The lowest $L_{\rm [CII]_{158\mu m}}/ L_{\rm [OI]_{146\mu m}}$ values are only found in AGNs, which indicate warm and dense gas components possibly heated by the AGNs (e.g., \citealt{herrera18a}).  
\qso{} exhibits a  $L_{\rm [CII]_{158\mu m}}/ L_{\rm [OI]_{146\mu m}}$ ratio of 6.5, which is among the lowest values found in local AGNs and smaller than local (U)LIRGs and dwarf galaxies. 
Similar $L_{\rm [CII]_{158\mu m}}/ L_{\rm [OI]_{146\mu m}}$ ratios are found in other high-$z$ systems (J1342+0928 and SPT 0418-47 on Figure \ref{diagnostics}(b), and J0439+1634 of 3.3, \citealt{yang19}) as well.
The  low $L_{\rm [CII]_{158\mu m}}/ L_{\rm [OI]_{146\mu m}}$ ratio measured from the nuclear region (within $\approx 3$ kpc) in \qso{} suggests a warm and dense neutral gas component, which is likely to be heated by the luminous AGN (see also Section \ref{firdeficit}). 
The $L_{\rm [OIII]_{88\mu m}}/ L_{\rm [NII]_{122\mu m}}$ ratio increases with increasing ionization parameter or decreasing metallicity (\citealt{nagao11}; \citealt{pereira17}; \citealt{{rigopoulou18}}).  The dust opacity could also influence the strengths of the $\rm [OIII]_{88\mu m}$ and $\rm [NII]_{122\mu m}$ lines (see Section \ref{metallicity}).
Local (U)LIRGs and AGNs reveal a wide range in the $L_{\rm [OIII]_{88\mu m}}/ L_{\rm [NII]_{122\mu m}}$ ratio, which may result from different ionization parameters in different galaxies (considering negligible differences in metallicity). 
The dwarf galaxy sample, however, shows systematically higher $L_{\rm [OIII]_{88\mu m}}/ L_{\rm [NII]_{122\mu m}}$ ratios compared to all local (U)LIRGs and AGNs, which can be attributed to the low metallicity in these dwarf galaxies. 
The $L_{\rm [OIII]_{88\mu m}}/ L_{\rm [NII]_{122\mu m}}$ ratio of 2.8 in \qso{} is within the range of local (U)LIRGs and AGNs, consistent with the value found in J1342+0928, and smaller than the $L_{\rm [OIII]_{88\mu m}}/ L_{\rm [NII]_{122\mu m}}$ ratio of 20.9 in SPT 0418-47.
A considerable fraction of the \nii{} emission can arise from the XDRs (see details in Sections \ref{hiicii} and \ref{metallicity}).
The comparable $L_{\rm [NII]_{122\mu m}/} L_{\rm FIR}$ ratio in \qso{} to the ratios found in local (U)LIRGs and AGNs, however disfavors a significant XDR contribution to the \nii{} line.
Accordingly, it is likely a result of (1) lower ionization parameter, (2) higher metallicity, or (3) higher dust opacity compared to SPT 0418-47 that causes the lower $L_{\rm [OIII]_{88\mu m}}/ L_{\rm [NII]_{122\mu m}}$ ratio in \qso{}.  This is also in agreement with the derived smaller metallicity in SPT 0418-47 than \qso{} (see Section  \ref{metallicity}).

In Figure \ref{diagnostics}(c), we show the $L_{\rm [CII]_{158\mu m}}/ L_{\rm [OI]_{146\mu m}}$ versus $L_{\rm [OIII]_{88\mu m}}/ L_{\rm [OI]_{146\mu m}}$ diagnostic diagram.
% \ion{O}{3} and  \ion{O}{2} are all oxygen bearing species thus the \oiii{}/ \oi{} ratio has negligible dependence on metallicity. The different gas phases that they trace enable the \oiii{}/ \oi{} ratio to probe the  volume filling factor ratio between the ionized and atomic gas. 
Because the \oiii{}/\oi{} ratio has negligible dependence on metallicity, it is a direct probe of the volume filling factor ratio between the ionized and atomic gas. 
 Local (U)LIRGs and AGNs have similar distributions of the $L_{\rm [OIII]_{88\mu m}}/ L_{\rm [OI]_{146\mu m}}$ ratio, while the high $L_{\rm [OIII]_{88\mu m}}/ L_{\rm [OI]_{146\mu m}}$ ratios in dwarf galaxies are due to the low gas density that enables UV photons to reach the outer regions of galaxies.
The high-redshift systems \qso{}, J1342+0928 and SPT 0418-47 reside within the range of local (U)LIRGs and AGNs.
 \qso{} exhibits a $L_{\rm [OIII]_{88\mu m}}/ L_{\rm [OI]_{146\mu m}}$ ratio of 1.9, which is over two times lower than J1342+0928 and SPT 0418-47, and approximately three times lower than the mean of the (U)LIRGs and AGNs, suggesting a slightly smaller ionized fraction in \qso{} compared to local (U)LIRGs, AGNs and other high-$z$ systems. This is also in agreement with the derived small contribution($\lesssim$17$\%$) of \cii{} from ionized gas in Section \ref{hiicii}.
Figure \ref{diagnostics}(d) shows the $L_{\rm [CII]_{158\mu m}}/ L_{\rm [OI]_{146\mu m}}$ versus $ L_{\rm [CII]_{158\mu m}}/ L_{\rm [OIII]_{88\mu m}}$ diagnostic diagram. 
Assuming that the majority of \cii{} emission arises from the neutral gas, the $ L_{\rm [CII]_{158\mu m}}/ L_{\rm [OIII]_{88\mu m}}$ ratio  could probe  the atomic and ionized gas volume filling factor ratio as well.
\qso{} reveals a $ L_{\rm [CII]_{158\mu m}}/ L_{\rm [OIII]_{88\mu m}}=3.4$ comparable with the mean of local (U)LIRGs and AGNs, and over five times of the values found in J1342+0928 and SPT 0418-47. 
%The slightly different result in comparing
That the $ L_{\rm [CII]_{158\mu m}}/ L_{\rm [OIII]_{88\mu m}}$ and $L_{\rm [OIII]_{88\mu m}}/ L_{\rm [OI]_{146\mu m}}$ ratios deliver different results for \qso{} compared to other samples is likely due to the additional dependence of the $ L_{\rm [CII]_{158\mu m}}/ L_{\rm [OIII]_{88\mu m}}$ ratio on metallicity and the variation of the \cii{} emission from ionized gas in different galaxies. Therefore, the $L_{\rm [OIII]_{88\mu m}}/ L_{\rm [OI]_{146\mu m}}$ ratio is a better tracer than the $L_{\rm [CII]_{158\mu m}}/ L_{\rm [OIII]_{88\mu m}}$ ratio when probing the atomic and ionized gas volume filling factor ratio.

%, which is in agreement with the results in Section \ref{hiicii} that $\geq 87\%$ of the \cii{} are from the neutral gas. The low (high) $L_{\rm [OIII]_{88\mu m}}/ L_{\rm [OI]_{146\mu m}}$ ($ L_{\rm [CII]_{158\mu m}}/ L_{\rm [OIII]_{88\mu m}}$ ) ratio in \qso{} is also in agreement the small contribution of the \cii{} from ionized gas region and the derived smaller ionized gas mass compared to the neutral gas mass.

\section{Summary}
In this paper, we report the detection of the \nii{} and \oi{} lines from the $z=6.003$ quasar \qso{}. 
Together with other FIR FS lines observed in this source, namely \cii{} and \oiii{}, \qso{} is among the $z\gtrsim 6$ quasars with the most complete set of FS lines available to date.
To probe the physical conditions of the atomic and ionized gas and evaluate the impact of AGN on the ISM, we compared the line emission to other measurements of local (U)LIRGs, AGNs and other high-redshift systems. The main results are summarized below.

$\bullet$ Of all the FS lines detected in \qso{}, the \cii{}, \nii{}, and \oiii{} lines show a ``line deficit'' relative to the FIR emission, which is comparable to local (U)LIRGs and AGNs. In contrast, the \oi{} line has a line-to-FIR flux ratio more than two times higher than that found in local galaxies, which suggests a warm and dense neutral gas component, that is possibly heated by the AGN.
%with $\textgreater 3 \times$ the $ L_{\rm [OI]_{146\mu m}/} L_{\rm IR}$ value found in local (U)LIRGs and AGNs  reveals no \oi{} deficit, suggesting a warm and dense neutral gas component or extra gas heating provided by the AGN. 

$\bullet$ From the observed \nii{} luminosity and the estimated \niis{} luminosity of \qso{}, we derive a density for the ionized gas of $\gtrsim$45 \pcubcm{} and estimate that only $\lesssim 17\%$ of the \cii{} emission originates in the ionized gas. This is in agreement with the results found in local (U)LIRGs, AGNs, and other high-redshift quasars.

 $\bullet$ We estimate a gas phase metallicity $Z/Z_{\sun}$=1.5$-$2.1 from the  \oiii{}/\nii{} ratio.

$\bullet$ Utilizing line ratios as ISM physical condition diagnostics, \qso{} exhibits a $L_{\rm [OIII]_{88\mu m}}/ L_{\rm [OI]_{146\mu m}}$ ratio more than two times lower than local (U)LIRGs, AGNs and high-redshift systems ($z=7.54$ quasar J1342+0928 and  $z=4.22$ lensed dusty star-forming galaxy SPT 0418-47),  suggesting a lower ionized gas fraction, consistent with the derived small contribution ($\lesssim 17\%$) of \cii{} line from ionized gas. 
\qso{} also reveals a $L_{\rm [OIII]_{88\mu m}}/ L_{\rm [NII]_{122\mu m}}$ ratio comparable to J1342+0928, local (U)LIRGs and AGNs, but more than seven times higher than SPT 0418-47, which is likely a result of a lower ionization parameter, a higher metallicity or a higher dust opacity for $\lambda_{rest} \lesssim 200 \mu m$.
Similar to other high-$z$ systems with available detections, \qso{} resides in the lowest $L_{\rm [CII]_{158\mu m}}/ L_{\rm [OI]_{146\mu m}}$ ratio region of local (U)LIRGs, indicative of warm and dense atomic gas most likely heated by the AGN.

Our work highlights future prospects for  \oi{} line detection by ALMA in other quasars and starburst galaxies  at $z \gtrsim 6$. Assuming a similar \cii{}/\oi{} ratio as that of J2310+1855, the $z \gtrsim 6$ quasars with a \cii{} flux $\gtrsim 2.0$ $\rm Jy\ kms^{-1}$ could be detected with S/N>3 in the \oi{} line with less than 35 minutes of ALMA time. 
As shown in this paper, the observations of a combination of various fine-structure lines in quasars and galaxies at $z \gtrsim 6$, including e.g., \oiii{}, \nii{}, \niis{}, \oi{}, \cii{} and \oione{} lines are critical in providing a detailed and unique diagnostic of the mutiphase ISM in these objects.

\section{acknowledgement}
This work was supported by the National Science Foundation of China (NSFC, 11721303, 11991052) and the National Key R\&D Program of China (2016YFA0400703). R.W. acknowledges supports from the NSFC grants No. 11533001 and the Thousand Youth Talents Program of China.
We thank T. Hashimoto, M. Pereira-Santaella, R. Herrera-Camus, and T. D{\'\i}az-Santos for kindly providing the data. 
We thank Feng Long for suggestions and help on the figures.
D.R. acknowledges support from the National Science Foundation under
grant numbers AST-1614213 and AST-1910107 and from the Alexander von
Humboldt Foundation through a Humboldt Research Fellowship for
Experienced Researchers.
FW acknowledges funding through ERC program Cosmic$\_$Gas.
YG's research is supported by National Key Basic Research and Development Program of China (grant No. 2017YFA0402704),
National Natural Science Foundation of China (grant Nos. 11861131007, 11420101002), and Chinese Academy of Sciences Key
Research Program of Frontier Sciences (grant No. QYZDJSSW-SLH008).
FB acknowledges support through the DFG Collaborative Research Centre 956.
This paper is based on ALMA observations: ADS/JAO.ALMA 2015.1.01265.S. ALMA is a partnership of ESO (representing its member states), NSF (USA) and NINS (Japan), together with NRC (Canada), MOST and ASIAA (Taiwan), and KASI (Republic of Korea), in cooperation with the Republic of Chile. The Joint ALMA Observatory is operated by ESO, AUI/NRAO and NAOJ.

\begin{deluxetable*}{cccccccccc}[p]
\tabletypesize{\tiny}
\tablecaption{Atomic and ionized fine-structure line observations and derived properties of \qso{}}
%\tablenum{2}
\tablecolumns{10}
\tablewidth{0pc}
\tablehead{
\colhead{Line}   &  \colhead{FWHM} & \colhead{S$\delta v$} &\colhead{Beam Size}   &  \colhead{Source Size}  &\colhead{Luminosity}&\colhead{$M_{gas}$}&\colhead{SFR} &\colhead{Ref.}\\
 \colhead{}  &  \colhead{(km s$^{-1}$)} & \colhead{(Jy $\rm km\ s^{-1}$)} &\colhead{(arcsec)}   &  \colhead{(arcsec)} & \colhead{($10^{9}\ L_{\sun}$)}&\colhead{ $\times 10^{7}$ $\rm M_{\sun}$}&\colhead{ $\rm M_{\sun}\ yr^{-1}$ } & \colhead{} \\
\colhead{(1)}&\colhead{(2)}&\colhead{(3)}&\colhead{(4)}&\colhead{(5)} &\colhead{(6)}&\colhead{(7)} &\colhead{(8)}&\colhead{(9)} }
\startdata
$\rm [NII]_{122um}$&328 $\pm$ 58&0.72 $\pm$ 0.20 &0.56 $\times$ 0.46, PA=-22\textdegree&(0.54 $\pm$ 0.19) $\times$ (0.31 $\pm$ 0.16), PA=105\textdegree$\pm$32\textdegree &  0.88 $\pm$ 0.24 &$\textgreater$62 &2208& L20  \\
$\rm [OI]_{146um}$&376 $\pm$ 16& 1.25 $\pm$ 0.08&0.58 $\times$ 0.47, \ PA=-1\textdegree &(0.50 $\pm$ 0.07) $\times$ (0.32 $\pm$ 0.09), PA=120\textdegree$\pm$20\textdegree &  1.28 $\pm$ 0.08 &$\textgreater$1700&11917& L20 \\
$\rm [OIII]_{88um}$&333  $\pm$  72&1.38  $\pm$  0.34&0.71 $\times$ 0.61, \ PA=-60\textdegree&(0.44  $\pm$  0.27) $\times$ (0.38  $\pm$  0.13), PA=70\textdegree$\pm$97\textdegree &2.44  $\pm$  0.61&$\textgreater$7.7 &1058& H19\\
$\rm [CII]_{158um}$&393 $\pm$ 21&8.83 $\pm$ 0.44&0.72 $\times$ 0.51, PA=10\textdegree&(0.55 $\pm$ 0.05) $\times$ (0.40 $\pm$ 0.07), PA=123\textdegree$\pm$44\textdegree &8.310 $\pm$ 0.414&$\textgreater$7800 &1524& W13
\enddata
\tablecomments{Column 1: line ID; Column 2 - 3: line width in FWHM and line flux. Note that the line flux is calculated through a single Gaussian fit to the line profile, and the flux uncertainty is the statistical error from the Gaussian fit; Column 4: beam size in FWHM, PA represents position angle; Column 5: source size deconvolved from the beam in FWHM; Column 6: line luminosity. Uncertanties are derived from the flux uncertainties;  Column 7: ionized (neutral) gas mass estimated from the $\rm [NII]_{122um}$ and $\rm [OIII]_{88um}$ ($\rm [OI]_{146um}$ and $\rm [CII]_{158um}$)  emission lines. Column 8:  SFR derived from the line luminosities. Column 9: references: This work (L20);  \cite{wang13} (W13); \cite{hashimoto18} (H19).}
\label{tabline}
\end{deluxetable*}

\begin{deluxetable*}{ccccc}[p]
\tabletypesize{\scriptsize}
%\tablenum{3}
\tablecaption{Continuum Properties}
\tablecolumns{8}
\tablewidth{0pc}
%\tablecaption{Continuum detection with AMLA}
%\tablecaption{Continuum properties}
\tablehead{
\colhead{Frequency}  & \colhead{S$\nu$} &\colhead{Rms}&\colhead{Beam Size}  & \colhead{Source Size}\\ 
\colhead{(GHz)}  & \colhead{(mJy)} & \colhead{($\rm \mu Jy\ beam^{-1}$)} &\colhead{(arcsec)}  & \colhead{(arcsec)} \\
\colhead{(1)}&\colhead{(2)}&\colhead{(3)}&\colhead{(4)}&\colhead{(5)} }
\startdata
344.2&14.94 $\pm$ 0.08& 49&0.55 $\times$ 0.44, PA=-27\textdegree& (0.29 $\pm$ 0.03) $\times$ (0.23 $\pm$ 0.02), PA=134\textdegree$\pm$16\textdegree\\
289.2&11.76 $\pm$ 0.04 &25&0.59 $\times$ 0.48, \  PA=-1\textdegree& (0.33 $\pm$ 0.02) $\times$ (0.25 $\pm$ 0.02), PA=154\textdegree$\pm$7\textdegree
\enddata
\tablecomments{Column 1: continuum frequency in observed frame; Column 2-3: continuum flux density and rms; Column 4-5: beam size and source size deconvolved from beam in FWHM, PA represents position angle. The deconvolved source sizes  of the emission lines (see Table \ref{tabline}) are $\textgreater 1.5 $ times larger than those derived from the continuum emissions. This is in agreement with the results found in ALMA observations of other high-z quasar host galaxies and  submillimeter galaxies (SMGs; e.g., \citealt{wang13,wang19}; \citealt{diaz16}; \citealt{venemans16, venemans17c}; \citealt{gullberg18}; \citealt{cooke18}; \citealt{rybak19}).  }
\label{tabcont}
\end{deluxetable*}

%spectra plural spectrum singular

\end{document}